\documentclass[pdflatex,sn-apa]{sn-jnl}

\usepackage{graphicx}%
\usepackage{multirow}%
\usepackage{amsmath,amssymb,amsfonts}%
\usepackage{amsthm}%
\usepackage{mathrsfs}%
\usepackage[title]{appendix}%
\usepackage{xcolor}%
\usepackage{textcomp}%
\usepackage{manyfoot}%
\usepackage{booktabs}%
\usepackage{algorithm}%
\usepackage{algorithmicx}%
\usepackage{algpseudocode}%
\usepackage{listings}%

\graphicspath{ {./pics/} }

\title[Epistemic integration and social segregation of AI in neuroscience]{Epistemic integration and social segregation of AI in neuroscience}

\author*[1]{\fnm{Sylvain} \sur{Fontaine}}\email{sylvain.fontaine@cnrs.fr}

\author[1]{\fnm{Floriana} \sur{Gargiulo}}

\author[1]{\fnm{Michel} \sur{Dubois}}

\author[2]{\fnm{Paola} \sur{Tubaro}}

\affil[1]{\orgdiv{GEMASS}, \orgname{CNRS-Sorbonne Universit\'e}, \orgaddress{\street{59-61 rue Pouchet}, \postcode{75017} \city{Paris}, \country{France}}}

\affil[2]{\orgdiv{CREST}, \orgname{CNRS-IPP}, \orgaddress{\street{5 avenue Henry Le Chatelier}, \postcode{91120} \city{Palaiseau}, \country{France}}}

\abstract{
In recent years, Artificial Intelligence (AI) shows a spectacular ability of insertion inside a variety of disciplines which use it for scientific advancements and which sometimes improve it for their conceptual and methodological needs.
According to the transverse science framework originally conceived by Shinn and Joerges, AI can be seen as an instrument which is progressively acquiring a universal character through its diffusion across science.
In this paper we address empirically one aspect of this diffusion, namely the penetration of AI into a specific field of research.
Taking neuroscience as a case study, we conduct a scientometric analysis of the development of AI in this field.
We especially study the temporal egocentric citation network around the articles included in this literature, their represented journals and their authors linked together by a temporal collaboration network.
We find that AI is driving the constitution of a particular disciplinary ecosystem in neuroscience which is distinct from other subfields, and which is gathering atypical scientific profiles who are coming from neuroscience or outside it.
Moreover we observe that this AI community in neuroscience is socially confined in a specific subspace of the neuroscience collaboration network, which also publishes in a small set of dedicated journals that are mostly active in AI research. 
According to these results, the diffusion of AI in a discipline such as neuroscience didn’t really challenge its disciplinary orientations but rather induced the constitution of a dedicated socio-cognitive environment inside this
field.
}


\begin{document}

\maketitle

\section{Introduction}

In recent years, Artificial Intelligence (AI) has been continuously spreading across science.
The associated worldwide scientific production and the amount of dedicated funding programs for technological developments supported both by academia and industry, are spectacularly growing \citep{gao_quantifying_2023,baruffaldi_identifying_2020,liu_tracking_2021}, and the outputs of such research are touching a variety of disciplines that are more and more citing it \citep{frank_evolution_2019,arencibia-jorge_evolution_2022}.
Though commonly associated to disciplines such as mathematics, statistics and computer science, AI originated as an interdisciplinary research area, and is currently spreading to a growing number of fields \citep{gargiulo_meso-scale_2023}.

AI knowledge and tools are amenable to applications in such welcoming disciplines which use them for scientific advancements, and which sometimes adapt and improve them in turn for their conceptual and methodological needs.
Bianchini et al. (\citeyear{bianchini_artificial_2022}) studied especially such mutual improvement dynamics at a macroscopic scale of science by focusing on the diffusion of neural networks' concepts and tools in numerous disciplines.
Qualified as a \textit{general method of invention} by these authors because of its non-disciplinary status, AI is thus expanding the \textit{adjacent possible} of the purposes of the concerned disciplines, and is therefore the source of innovation in the latter \citep{kauffman_investigations_2000, monechi_waves_2017}.

In this paper we address the process of penetration of AI inside a given field of research. 
We focus on neuroscience, which are part of such fields largely receiving AI, according to Gargiulo et al. (\citeyear{gargiulo_meso-scale_2023}).
We choose especially this field as it has a particular history which is closely intertwined with the AI one.
Indeed neuroscience and AI are continuously providing feed-backs to each other toward the comprehension of the human brain and the artificial reproduction of some cognitive processes \citep{hassabis_neuroscience-inspired_2017, savage_how_2019}.
This lead us to also assume that neuroscience is not only a filed that receives AI, but also contribute to reshape AI knowledge, tools and practices, thus expanding the \textit{adjacent possible} of the latter.

Along these lines, we focus on two dynamical mechanisms underlying the appropriation of AI by neuroscience and then the possible knowledge transfer between the two fields:
\begin{itemize}
    \item The cognitive embedding of AI into the multidisciplinary context of neuroscience: 
    \begin{enumerate}
        \item what is the disciplinary landscape around neuroscience research which is using AI?
        \item how does such research fit in the disciplinary objectives of neuroscience?
    \end{enumerate}
\item The diffusion throughout the neuroscience community:
    \begin{enumerate}
        \item[3.] who are the actors leading AI research in this field?
        \item[4.] does AI-related knowledge widespread throughout the whole scientific community?
    \end{enumerate}
\end{itemize}

In what follows we answer these questions by analyzing a large corpus of neuroscience articles published between 1970 and 2019, extracted with a journal-oriented query from the Microsoft Academic Knowledge Graph (MAG) \citep{ghidini_microsoft_2019}.
We distinguish AI-related publications from others with a dedicated keyword filter applied to their titles and abstracts, as in \citep{gargiulo_meso-scale_2023}.
After discussing in Section~\ref{sec:literature_review} the intertwined history of AI and neuroscience, and also the main concepts mobilized in this paper, we detail in Section~\ref{sec:data} the building of some relational structures and some metrics given by the corpus that will be analyzed afterwards, namely its temporal egocentric citation network (composed of bibliographical references and citations of each paper), its peer-reviewed journals, and its temporal collaboration network, featured with the disciplinary profiles of the involved scientists.

In Sections~\ref{sec:results_disc_jaccard} and \ref{sec:results_disc_carto}, we identify within the citation network the disciplines that are shaping the AI research in neuroscience, and the extent to which they differ from those that are characterizing other neuroscience's subfields unrelated to AI.
We especially show that some disciplines related to STEM have become the main influence of AI research since the 1990s, by replacing progressively the conventionally used ones in neuroscience and commonly associated with biomedical and clinical research.
The disciplines impacted by AI research over time are also shifting from neuroscience-related clinical and biomedical fields to STEM ones.
These first results thus show that AI research is building itself as a subfield (or specialty) inside neuroscience, which draws upon a disciplinary basis that is different from the other neuroscience subfields.
We also confirm this partial cognitive integration in Section~\ref{sec:results_journals}, which reveals the spreading of AI in almost all the journals of the database, but with different patterns of promotion of such research.
In particular, we show that the journals that are publishing a lot of AI-related works are invested by researchers who contributed to AI research in general (inside or outside neuroscience).
By regarding the temporal co-authorship network of the field, we show in the following Section~\ref{sec:results_authors} that the AI community includes two main academic profiles: (1) researchers trained in neuroscience -- mainly within the biomedical and clinical parts --, who publish few AI-related works, and (2) researchers trained in STEM-related disciplines and having moved to neuroscience, who contribute much more to AI research in the welcoming field than those described by profile (1).
Putting all the AI practitioners together, we then show that they are more and more isolated in the collaboration network since 1970.

Finally, by considering AI as a \textit{research-technology} spreading across science, as originally conceived by Shinn and Joerges (\citeyear{shinn_transverse_2002}) and formalized later by Hentschel (\citeyear{hentschel_periodization_2015}), we discuss in Section~\ref{sec:discussion} the partial diffusion observed empirically in our results.
With this framework we also discuss the similarity and differences between the local development dynamics of AI within this particular field of research, and the global one within the whole science system that is depicted in \citep{gargiulo_meso-scale_2023}.

\section{Literature review}
\label{sec:literature_review}

\subsection{AI and neuroscience: toward the mutual expansion of their \textit{adjacent possible}}

Artificial Intelligence (AI) commonly refers to both a research program and, more generally, a set of complex computer-based programs which aim to mimic human mind processes with high reckoning power.
Although its foundations are mainly associated with STEM disciplines, mainly mathematics, statistics and computer science \citep{gargiulo_meso-scale_2023}, AI is constantly evolving alongside neuroscience by maintaining a virtual circle of mutual improvement \citep{hassabis_neuroscience-inspired_2017}.

Indeed neuroscience brings in a first place empirical confirmations of some theoretical models that reproduce parts of mental processes, and that were first imagined, analytically derived and computationally simulated by (neuro)psychologists within the field of cognitive science \citep{cooper_cognitive_2010, lake_building_2017}.
Most of these models are at the roots of AI-related algorithms, notably the case of the bio-inspired AIs such as artificial neural networks and their numerous versions, which became \textit{biologically plausible} with neuroscience.
In particular, the 1980s and early 1990s marked also the launching of the first body-scanner machines applying positron emission tomography (PET) and functional magnetic resonance imaging (fMRI), both of which led to important discoveries on the functional biological mechanisms in the human brain that are induced by complex cognitive tasks \citep{cooper_cognitive_2010}.
This contributed to the rise of the connectionist paradigm that is now dominant in brain sciences, even if some debates are persisting within cognitive science about the representations of knowledge and the logical operations to process them in interaction with the real world \citep{mccarthy_epistemological_1981, andler_connexionnisme_1990, perconti_deep_2020}

In the recent \textit{big data} era, a panel of AI tools enables the efficient processing of large datasets composed of various kinds of biomedical data (electroencephalogram, MRI, biomarkers tracking, movement recordings, psychological survey, etc.), acquired from important clinical trials and cohorts for studying brain damages \citep{gopinath_artificial_2023}.
They are especially part of the improvement of the diagnosis of various neuro-degenerative diseases and of the attribution of potential dedicated treatments, if they exist.

This virtual feedback loop, well documented in the neuroscience literature, is thus inducing a reinforcement dynamics of both AI and neuroscience which are receiving it. 
From this assertion we assume that AI and neuroscience are expanding each other's \textit{adjacent possible} \citep{kauffman_investigations_2000, monechi_waves_2017}, ie. one domain is extending the field of possibilities that have yet to be explored in the other, by blending with the pre-existing knowledge and practices that are characterizing the latter.
One domain is thus reshaping the knowledge space of the other.

\subsection{Conceiving AI as a \textit{research-technology} in science}

In a socio-historical perspective of science, Shinn and Joerges (\citeyear{shinn_transverse_2002}) propose the notion of \textit{research-technology}, called also \textit{transverse science}, to describe the dynamics of science since the end of the $19^{th}$ century, now largely based on instrumentation for experimental or empirical investigations, especially in physical and life sciences.
Within such a research regime, the production of knowledge is conditioned by an instrument that is designed in a specific research environment before being disseminated outside the latter \citep{shinn_transverse_2002, marcovich_science_2020}.
Such an instrument requires the contribution of a dedicated socio-cognitive workforce composed of a variety of actors (scientists, technicians, promoters,  administrators,...), to develop a dedicated technological culture of it, ie. a set of associated knowledge and practices shared by everyone in the community, without necessarily claiming a common professional identity related to it.
These criteria shape a social group that could be also defined as an \textit{epistemic community} \citep{haas_introduction_1992,roth_reseaux_2008}, although the research-technology framework relaxes the socio-cognitive boundaries that are specific to such communities.
According to Shinn and Joerges (\citeyear{shinn_transverse_2002}), the actors mentioned above are indeed able to move between different research environments and across the established disciplinary boundaries in order to provide their expertise for the resolution of diverse scientific problems.

The research-technology associated with an instrument is thus a dynamical entity, with changing social and institutional organizations at different moments of its development.
Based on Shinn and Joerges (\citeyear{shinn_transverse_2002}), Hentschel (\citeyear{hentschel_periodization_2015}) proposes four stages to describe the life of a given research-technology in a historical perspective, ie. throughout a long time period: \textit{prehistory}, \textit{exploration}, \textit{optimization} and \textit{diffusion}.

The first two stages are often associated together in the designing and testing phases of a given instrument.
According to these authors, the underlying process needs temporarily a closure of the community working on it, thus fostering the creation of a dedicated scientific field, often interdisciplinary, in which the instrument is also a research object.

Shinn and Joerges (\citeyear{shinn_transverse_2002}) also highlight the criterion of \textit{genericity} of an instrument: at later stages of its development, it could be adapted for disciplinary research contexts other than those in which it has been originally designed, or for a variety of applications outside scientific research.
The two last development steps of Hentschel are thus representing such dynamics toward this final step.
In particular, the diffusion phase requires a relaxation of the disciplinary boundaries within which the instrument was designed during the prehistoric and exploratory phase, thus implying an openness of the actors in the community, who promote it in different fields of research within academia or industry.
This the case of technologies related to laser beams or X-ray emission, now use widely from experimental physics to medical research and standard practices, even in daily objects such as smartphones \citep{hentschel_periodization_2015}.

According to the analysis of the AI research ecosystem provided by Gargiulo et al. (\citeyear{gargiulo_meso-scale_2023}), AI seems to embrace some characteristics of an instrument embedded in a global research-technology program within science.
Through the analysis of the development of AI in neuroscience, throughout a longer time frame (1970-2019), we discuss in Section~\ref{sec:discussion} to what extent this framework applies in the case of the development and the diffusion of AI in a single field of research.

\section{Data and methods}
\label{sec:data}

\subsection{Extraction and preprocessing of the data}
\label{sec:build_dataset}

MAG is a scientometric database that is well used in \textit{science of science}, and especially for drawing the main trends of research within AI research \citep{gargiulo_meso-scale_2023}, measuring their impact in different scientific disciplines \citep{gao_quantifying_2023}, identifying the main actors involved in such research \citep{frank_evolution_2019} and how they are organized accross the world \citep{tang_internationalizing_2022}, among others.
Much bigger than the traditional databases such as the \textit{Web of Science} (WOS) and \textit{Scopus} \citep{visser_comparison_2021}, MAG covers not only the articles published in peer-reviewed journals, but also the publications within conference proceedings, books, preprints, etc., for a total of about 260 million scientific publications in all disciplines (in 2022).

MAG is organized as a big network of files containing different metadata to describe a single publication, such as title, abstract, publication date and publisher \citep{ghidini_microsoft_2019}. 
These metadata are embedded as strings or numerical identifiers in a set of dump files linked together by link files, all of them being available on the data deposit platform Zenodo \citep{mag_2020}.

To build an exhaustive database of the neuroscience literature from MAG, we draw upon a list of journals that we extracted from WOS, with the help of the database \textit{SCImago Journal Rank} (SJR).
We choose to use SJR in addition of WOS in order to extend the scope of neuroscience too narrow in the classification of the last one.
Indeed the former is covering more neuroscience journals than the latter: in 2021, WOS referenced only 281 journals labeled as \textit{Neuroscience} while SJR referenced 608 journals labeled as such.
Moreover journals could be labeled differently by the two databases.
Thus, with the help of SJR, we can retrieve in WOS some journals that are not labeled as \textit{Neuroscience} in this database but that are in fact publishing papers treating neuroscience topics.
For instance, the well-recognized journal Neurocomputing, while labeled with both \textit{Neuroscience} and \textit{Computer Science} in SJR, is labeled only as \textit{Computer Science} by WOS.
Using SJR thus contributes to increasing the diversity of neuroscience-related subjects treated by the articles in the final dataset under study.

After having extracted the set of journals labeled as \textit{Neuroscience} by WOS, we add the set of journals also labeled as such by SJR, and remove duplicates occurring in both of them.
We thus get 421 journals.

Here we focus only on peer-reviewed journals because of their easy availability in WOS database, although we are aware that AI research could be largely published in other publication media such as preprint archives and conference proceedings \citep{wainer_how_2013}.

Then we extract the publications from MAG with an ISSN identifier matching procedure based on this list of journals.
These publications are brought together in a set denoted by $\mathcal{P}$ in the following.
In order to distinguish AI-related works in our neuroscience dataset, we apply a selection criterion such that they must include at least one AI-related keyword in their title or abstract (see the SI of Gargiulo et al. (\citeyear{gargiulo_meso-scale_2023}) for the complete list of keywords).
This subset is denoted $\mathcal{P}\cap AI$.
All neuroscientific studies outside this set are called $\mathcal{P}\cap\overline{AI}$.

\begin{figure}[htb]
    \centering
    \includegraphics[width=0.9\textwidth]{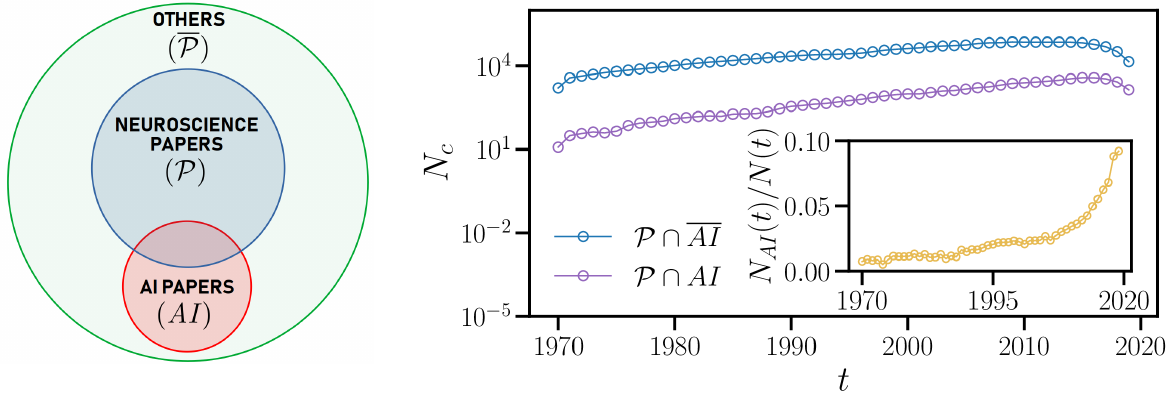}
    \caption{Left: Classification of the papers in the extracted corpus. 
    The intersection between the blue and red zones corresponds to the set $\mathcal{P}\cap AI$ in the main text, while the blue zone excluded from this intersection represents the set $\mathcal{P}\cap\overline{AI}$. 
    The green set $\overline{\mathcal{P}}$ includes all papers added after the building of the ego-centered citation network that are not published in neuroscience journal.
    Right: Cumulative number $N_c$ of publications in the two main subsets considered in this paper. 
    The inset shows the instantaneous part of AI-related publications in the whole neuroscience corpus.}
    \label{fig:class_papers+cumul_nb_papers}
\end{figure}

In the following we keep only the papers that were published in the period running from 1970 until 2019, and that have at least 10 references and at least 10 generated citations.
Finally the sets $\mathcal{P}\cap\overline{AI}$ and $\mathcal{P}\cap AI$ include respectively 829,317 and 26,374 papers, spread over the period mentioned as illustrated in Fig.~\ref{fig:class_papers+cumul_nb_papers}.
Therefore, among all this research in neuroscience, only 3\% contain AI-related keywords.
The inset of this figure exhibits especially a slow growth of the share of AI-related publications in neuroscience at the very end of the 1980s, stabilizing around 1995 and followed by a very rapid growth from 2007 until 2019.
This plateau around 1995 suggests a second, prolonged \textit{AI winter} in neuroscience, characterized here by stable interest in AI research but not waning, unlike in other disciplines or research fields \citep{cardon_neurons_2018,schuchmann_history_2019}.
The following period of important growth of this share, started in 2007, suggests also the main influence of the rise of deep learning techniques, a well-known trend shared by almost all the sciences \citep{cardon_neurons_2018,gargiulo_meso-scale_2023}.
The share of AI publications in neuroscience reaches only 10\% of the number of publications at its highest stage situated at the end of the studied period, which means that the use of the AI keywords of our list in neuroscience remains rather limited, even today.

\subsection{Capturing the disciplinary landscape}
\label{sec:build_disciplinary_landscape}

\subsubsection{Building of a citation ecosystem}
\label{sec:build_citation_ecosystem}

\begin{figure}[htb]
    \centering
    \includegraphics[width=0.7\textwidth]{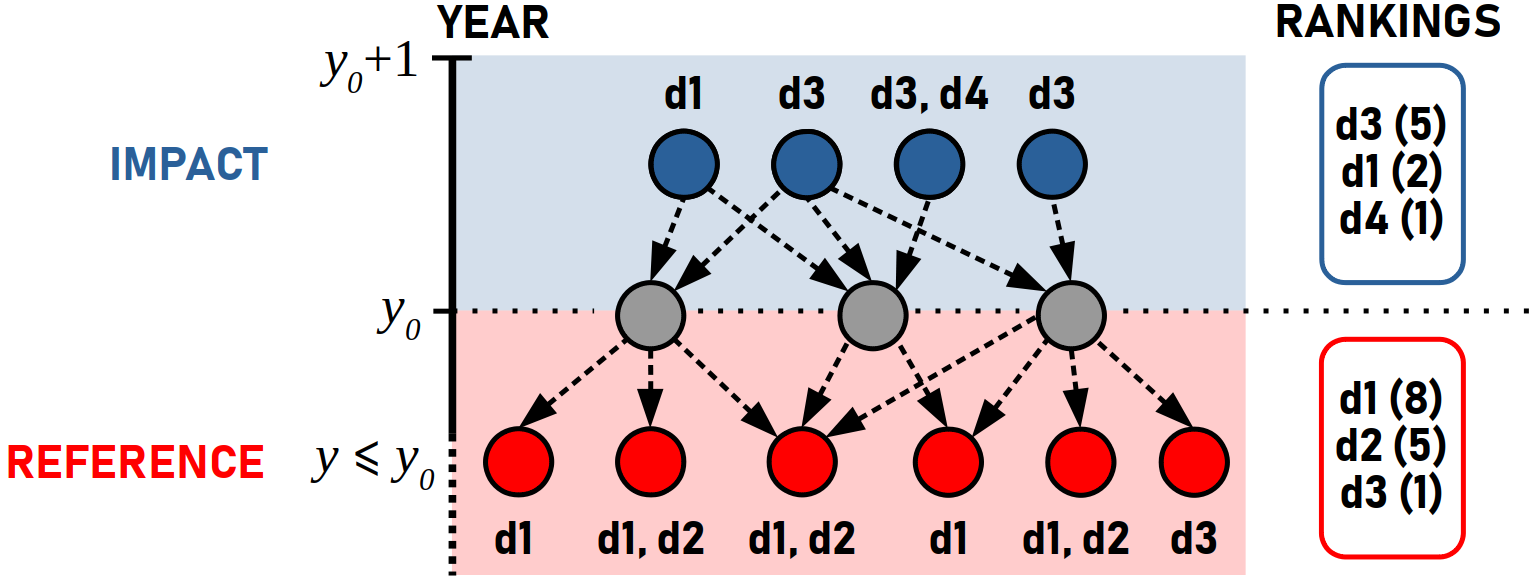}
    \caption{Citation ecosystem centered here around three papers (grey) published at year $y_0$ and that are included either in $\mathcal{P}\cap AI$ or $\mathcal{P}\cap\overline{AI}$.
    One dashed arrow represents the citation of a target paper by a source one.
    Hence red points are the papers that are \textit{cited} by the papers of our corpus (\textit{reference}) while the blue ones are \textit{citing} them (\textit{impact}).
    The rankings are shown in decreasing order with the associated number of citations of each JSC $d_i$.}
    \label{fig:ecosystem}
\end{figure}

After gathering the dataset under study, we build an exhaustive citation ecosystem around neuroscience publications, as illustrated in Fig.~\ref{fig:ecosystem}.
We use the general reference link file of MAG -- which is actually a giant citation network -- for retrieving the references and citations of our identified papers.
We thus add such papers citing or cited in those in our dataset, as long as they are published in a journal that belongs to the WOS journals database -- we conserve here 84\% of the original citation network centered on the neuroscience dataset $\mathcal{P}$.
The added papers could therefore be published either in neuroscience publications, therefore in the set $\mathcal{P}$, or in others such that they join the set denoted by $\overline{\mathcal{P}}$, ie. the green border zone in Fig.~\ref{fig:class_papers+cumul_nb_papers}.
With the journal classification of WOS, we then enrich these sets of references and citations by assigning them a set of Journal Subject Categories (JSC) associated to their journals if they are referenced in this database, which is a synonym of a large well-recognized discipline in science.

Then, for each corpus $\mathcal{P}\cap AI$ and $\mathcal{P}\cap\overline{AI}$, we count the annual number of citations obtained by each JSCs represented in the global set of bibliographic references used by their papers.
Doing this year by year over the period 1970-2019, we therefore build a temporal ranking of fields of study that are most cited by each corpus.
For example, in Fig.~\ref{fig:ecosystem}, the papers cited by one corpus at year $y_0$ (grey points) are mainly published in journals labeled by the JSC $d_1$ with 8 citations, followed by the JSCs $d_2$ with 5 citations and $d_3$ with only one (red points and corresponding red-contoured ranking).
Hence the disciplinary reference ranking $r(y_0)=\{d_1,d_2,d_3\}$ with each JSC sorted according to their respective ranks mentioned above.
However this ranking could be different before this year $y_0$ and also in the following ones, depending on the citations received by each JSC at these years.

We do the same for the JSCs that are most citing our papers of interest up to one year after their official publication year.
We choose this temporal period to avoid the high time dispersion of the generated citations in the studied temporal period, the citation impact by the papers published in 1970 being eventually higher than those published in 2019.

In this way we assess from which JSCs are issued at first the generated citations before eventually spreading into a broader disciplinary landscape.
Finally, by looking at both reference and impact side, we represent two disciplinary landscapes associated respectively with the corpora $\mathcal{P}\cap\overline{AI}$ and $\mathcal{P}\cap AI$.

\subsubsection{Representation of the disciplinary landscape}
\label{sec:build_2D_maps}

Following the procedure described in Section~\ref{sec:build_citation_ecosystem}, and according to all the couples of references' and citations' numbers associated with all the represented JSCs in $\mathcal{P}\cap \overline{AI}$ and $\mathcal{P}\cap AI$, we associate with each JSC $d$ two pairs of ranks indicating its weight either in references or in citations of the two corpora at year $t$, denoted respectively as $r^d_R(t)=\left(r^d_{R,\mathcal{P}\cap \overline{AI}}(t), r^d_{R,\mathcal{P}\cap AI}(t)\right)$ and $r^d_I(t)=\left(r^d_{I,\mathcal{P}\cap \overline{AI}}(t), r^d_{I,\mathcal{P}\cap AI}(t)\right)$.
In case of a missing JSC in one of the two corpora at a given year $t$, we fill the missing rank by a maximum value set at 100.

\begin{figure}[htb]
    \centering
    \includegraphics[width=0.35\textwidth]{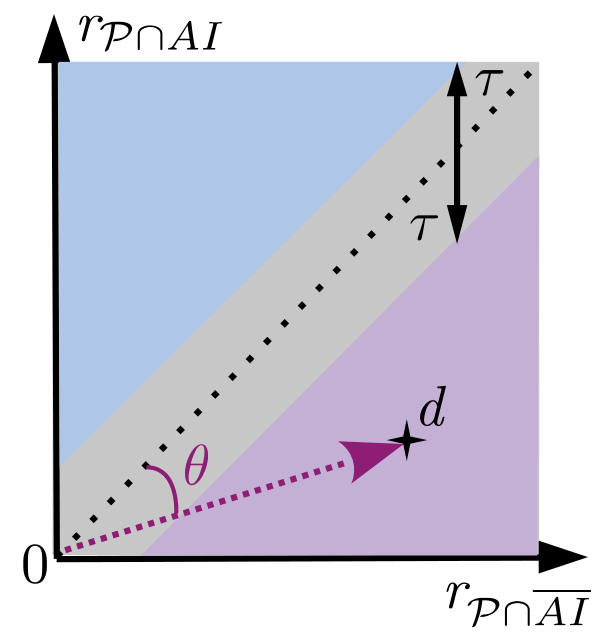}
    \caption{Coordinate system to characterize the distribution of disciplines in the references and citations of the corpora $\mathcal{P}\cap \overline{AI}$ and $\mathcal{P}\cap AI$.}
    \label{fig:scheme_rankings_angles}
\end{figure}

With one of such pair of ranks (either for references or citations), we can locate one discipline $d$ in a 2D space of rankings, as shown in Fig.~\ref{fig:scheme_rankings_angles}.
In this coordinate system, the lower the value of one axis, the better the associated rank and the higher the number of collected citations along this axis.
The angle $\theta$ indicates the deviation of $d$ from the diagonal in this space, where the ranks are exactly the same in each corpus.

From this map we define the \textit{common interest area} between $\mathcal{P}\cap AI$ and $\mathcal{P}\cap \overline{AI}$ (in grey) as the region close to the diagonal where the two rankings of the fields of study inside it are almost the same and not significantly varying over time.
This zone is comprised between the two lines of respective equations $r_{\mathcal{P}\cap AI} = r_{\mathcal{P}\cap\overline{AI}}+\tau$ (above the diagonal) and $r_{\mathcal{P}\cap AI} = r_{\mathcal{P}\cap\overline{AI}}-\tau$ (below the diagonal), with $\tau$ a parameter set as 10 in this paper.

The blue zone over the common interest area corresponds to the space where the disciplines have a better rank in $\mathcal{P}\cap \overline{AI}$ than in $\mathcal{P}\cap AI$.
In the case of references, such disciplines in this zone are more cited by the corpus $\mathcal{P}\cap \overline{AI}$ than by $\mathcal{P}\cap AI$, and in the case of citations received by each of the two corpora, such disciplines in this zone cite more the papers in $\mathcal{P}\cap \overline{AI}$ than those in $\mathcal{P}\cap AI$.
The purple zone under the common interest area is therefore the opposite case, ie. in the case of references (for example), the disciplines in this zone are more cited by the corpus $\mathcal{P}\cap AI$  than by $\mathcal{P}\cap \overline{AI}$.

In Section~\ref{sec:results_disc_carto}, we apply this per decade periods instead of single years, ie. we build a decennial ranking space from the total number of citations given or received by the represented JSCs in $\mathcal{P}\cap \overline{AI}$ and $\mathcal{P}\cap AI$ within the period under study.

\subsection{Capture use of AI by neuroscientists}

\subsubsection{Classification of authors according to their AI involvement}
\label{sec:build_authors_dataset}

The dataset $\mathcal{P}$ alone would not be sufficient for a study of social dynamics of scientists inside neuroscience.
Indeed the latter would have had published in disciplines or fields of research other than neuroscience or in journals that are not labeled as such in WOS.
We thus add another level of granularity by extracting all the missing publications that have been published by the authors who belong to $\mathcal{P}$.
Notice that some of these publications can have been published in the neuroscience journals extracted in our list of 421 journals (see Section~\ref{sec:build_dataset}), since we focus only on the more impacting authors who have published papers with at least 10 generated citations (and having at least 10 references).
From this extended set of papers we deduce the duration of the scientific life of each author, and we select first those who began after 1940.
In addition, because of the poor disambiguation of some authors in the provided MAG dump-files, where some names are associated with centuries-long career and far too many publications, we select only the authors having a reasonable duration of academic life, here up to 50 years.

\begin{figure}[htb]
    \centering
    \includegraphics[width=0.5\textwidth]{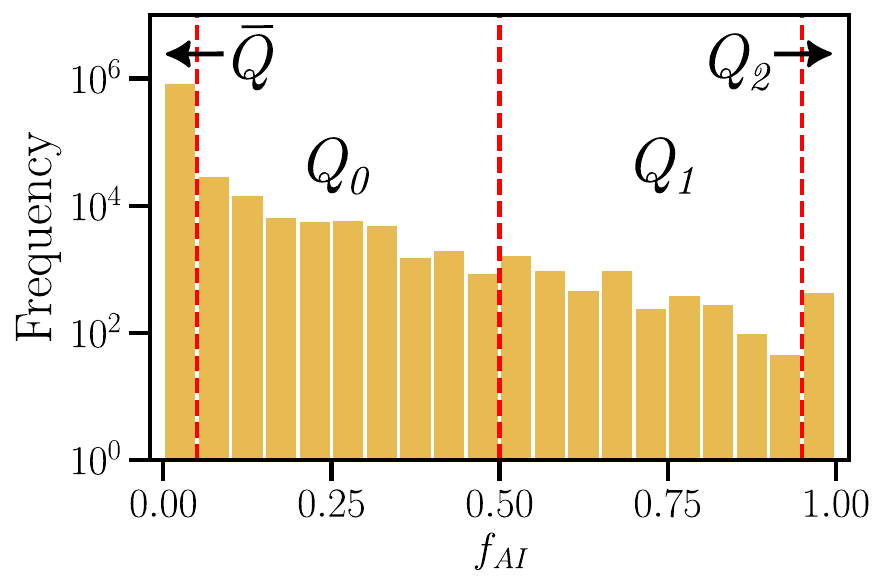}
    \caption{Distribution of the AI score $f_{AI}$ of the authors.}
    \label{fig:AIscore_distrib}
\end{figure}

With these added data, we also capture the real activity (before 2019) of each author in AI research, not only in neuroscience but also in other fields of research by attributing them an AI score.
We define it as the share of AI-related papers published before 2019, $f_{AI}(a) = n^{\mathcal{P}\cap AI}_a/n^{tot}_a$.
To prevent the unexpected effect of accumulation of scientists having an AI score equal to 1 due to a single publication in their very short career, we filter a second time the authors dataset by considering only those who have published at least 3 papers.
We thus consider in the following 886,074 scientists, in which 188,325 (only 16\% of the scientists) have published at least one AI-related paper in neuroscience.
According to the $f_{AI}$ distribution shown in Fig.~\ref{fig:AIscore_distrib}, we divide this set of authors in four parts, namely $\overline{Q}$ ($f_{AI}=0$; $N_a=697749$), $Q_0$ ($f_{AI}\in(0,0.5)$; $N_a=182925$), $Q_1$ ($f_{AI}\in[0.5,1)$; $N_a=4977$) and $Q_2$ ($f_{AI}=1$; $N_a=423$).

\subsubsection{Collaboration network}
\label{sec:build_collaboration_network}

We then build a temporal co-signature network TCN from the set of papers $\mathcal{P}$.
Two authors $i$ and $j$ publishing together at year $t$ are linked by an edge weighted by the number of common publications at this year $w_{ij}(t)$.
All co-signatures in all papers published in year $t$ shape a weighted snapshot called TCN($t$).
In order to unveil the structure of the collaboration at macro-scale in the whole period 1970-2019, we build the weighted time-aggregated co-signature network ACN, which includes the set $V$ of all authors appearing in the dataset, and a set of weighted edges obtained by summing all the weights of the edges appearing in each snapshot of TCN, $E=\lbrace \tilde{w}_{ij} \, \forall (i,j)\in V^2 | \tilde{w}_{ij} = \sum_t w_{ij}(t) \rbrace$.
ACN is finally composed of 871,282 nodes and 7,420,423 weighted edges, and is divided into 17,599 independent connected components, with a giant one including 93\% of the nodes.
The smallest components exhibit also a non-negligible representation of authors belonging to the quartiles $Q_1$ and $Q_2$, respectively 25,5\% and 26,5\% of the population in these quartiles.
Since the authors belonging to them are supposed to drive the AI research in neuroscience, we choose to keep all the communities for our analysis, even smaller than the giant component.

Here the collaboration network is built only among the most impacting scientists of our dataset.
Indeed we draw on the one hand the collaboration network directly from the set of neuroscience papers that are the most impacting in terms of citations, and on the other hand we focus on the authors who published at least 3 articles up to 2019.
Therefore we disregard the short-terms collaborations among very early-career scientists, short-career scientists or extra-academic ones, who are involved in one or two publications with too small impact but who can potentially drive forward innovation.

\subsubsection{Intensity of collaborations between $f_{AI}$ quartiles}
\label{sec:z-score_collab}

In this section we detail the general computation of z-scores which will be used in Section~\ref{sec:results_authors_collab} to indicate the propensity of collaboration in the time-aggregated collaboration network ACN.

We first consider a static collaboration network $G=\{S,E\}$ with $S$ the set of scientists and $E$ the set of edges between them.
Each scientist in $S$ is identified by their respective AI score $f_{AI}$.
According to Section~\ref{sec:build_authors_dataset}, the distribution is divided in four parts $\{\overline{Q},Q_0,Q_1,Q_2\}$, from the least to the most AI-expert.
Therefore one author is belonging to one given group $Q$ given its own AI score.
By aggregating the scientists into such $Q$ groups, we compute the $4\times4$ matrix $\tau_{obs}$ of effective edges between these groups, where one element is written as $\tau_{obs}(Q_i,Q_j)=|\{(u,v)|u\in Q_i, v\in Q_j\}|$.

Then we create a set of $N$ alternative collaboration networks $\{\tilde{G}_k\}=\{(S,\tilde{E}_k)\}$, with $k=\{1,...,N\}$, which are all based on the same set of scientists but have different sets of edges obtained by perturbing the network with a uniform shuffling which conserves the degree distribution of the scientists.
By applying the same clustering of scientists in $Q$ groups for each network as for the real network $G$, we draw a set of matrices $\{\tau_k\}$ corresponding to the share of edges between the $Q$ groups within the randomized networks  $\{\tilde{G}_k\}$.
From this set we extract on the one hand the average matrix $\tau_{sim}$ where one element is defined as $\tau_{sim}(Q_i,Q_j)=\frac{1}{N}\sum_k\tau_k(Q_i,Q_j)$, and on the other hand the standard deviation matrix $\sigma$ where one element is defined as $\sigma(Q_i,Q_j)=\frac{1}{N-1}\sum_k(\tau_k(Q_i,Q_j)-\tau_{sim}(Q_i,Q_j))^2$.

We finally compare the empirical matrix $\tau_{obs}$ of the real network with the simulated one $\tau_{sim}$ through a $z$-score matrix where one element is defined as follows:
\begin{equation}
    z(Q_i,Q_j) = \frac{\tau_{obs}(Q_i,Q_j) - \tau_{sim}(Q_i,Q_j)}{\sigma(Q_i,Q_j)}\,.
    \label{eq:z-score_qqbar}
\end{equation}

This standardization is used here to test the over-representation or under-representation of a given number of edges between two groups with respect to the corresponding average simulated value which represents an ideal situation through randomization.
Applying the whole procedure before on the network ACN built in Section~\ref{sec:build_collaboration_network}, with $N=100$, gives the Fig.~\ref{fig:collabNet_quartile}B.
Fig.~\ref{fig:collabNet_quartile}A is also drawn by applying the same method with same $N$ on each temporal snapshot of the temporal collaboration network TCN (see also Section~\ref{sec:build_collaboration_network}), and by aggregating the groups $Q_0$, $Q_1$ and $Q_2$ into a single one called $Q_i$ (see Fig.~\ref{fig:collabNet_quartile}A) that includes all the scientists with at least one AI-related publication.
Since TCN is undirected, the $4\times4$ matrices $\tau_{obs}$, $\tau_{sim}$, $\sigma$ and $z$ are therefore reduced to scalars defined with only the two groups $\overline{Q}$ and $Q_i$.

\section{Results}
\label{sec:results}

\subsection{Citation homogenisation of the AI research specialty with general neuroscience}
\label{sec:results_disc_jaccard}

First of all we explore the integration of AI technologies and knowledge within neuroscience through the dynamical interaction between the respective disciplinary environment of the two fields -- AI and neuroscience.
We especially compare year by year the basis of reference and citations that are shaping these fields, that are here represented as the AI-related publications and the non-AI ones within our neuroscience dataset.

After having divided the neuroscientific papers in two sub-corpora, one including the AI-related papers ($\mathcal{P}\cap AI$) and another including non-AI ones ($\mathcal{P}\cap \overline{AI}$), we build for each of them two disciplinary rankings which could vary over time, one for the disciplines appearing in their respective references ($r^R(t)$), called \textit{reference ranking} in the following, and another for those appearing in the respective citations they have received ($r^I(t)$), called \textit{citation ranking} in the following.
The reference ranking associated with one given corpus summarizes its main influential fields of study, thus the disciplinary structure on which it draws over time, while the citation ranking indicates the fields that are impacted at first by the corpus (for more detail, see Section~\ref{sec:build_citation_ecosystem} above).

We then compare macroscopically at each year the reference rankings of the two corpora ($r^R_{\mathcal{P}\cap AI}(t)$ and $r^R_{\mathcal{P}\cap\overline{AI}}(t)$) and also their citation rankings ($r^I_{\mathcal{P}\cap AI}(t)$ and $r^I_{\mathcal{P}\cap\overline{AI}}(t)$) by using a common similarity metric $J$ provided by Gargiulo et al. (\citeyear{gargiulo_classical_2016}, see their supplementary materials).
This last measure, comprised between 0 and 1, evaluates how similar are two given rankings $r_A$ and $r_B$: if $J(r_A,r_B)=1$, the two rankings are exactly the same, ie. containing the same elements with the same respective ranks; conversely, if $J(r_A,r_B)=0$, the elements included in $r_A$ are not in $r_B$, whatever their respective ranks.

\begin{figure}[htb!]
    \centering
    \includegraphics[width=0.7\textwidth]{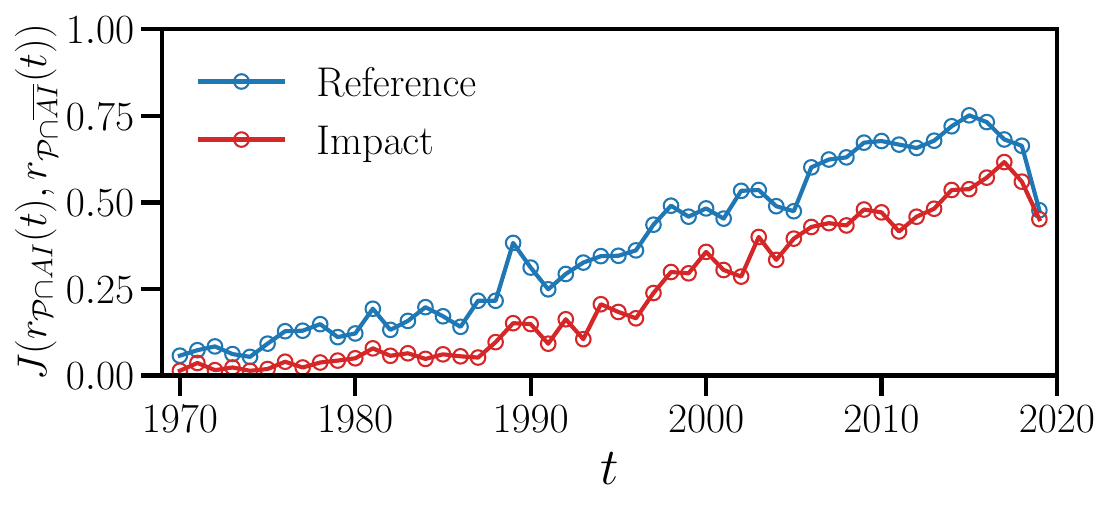}
    \caption{Instantaneous similarity between the references' or citations' (also called impact) rankings of the two corpora $\mathcal{P}\cap AI$ and $\mathcal{P}\cap\overline{AI}$ at year $t$.}
\label{fig:jaccard}
\end{figure}

Fig.~\ref{fig:jaccard} shows the evolution of two such similarity measures, a first one for comparing the rankings of disciplines cited by the corpora $\mathcal{P}\cap AI$ and $\mathcal{P}\cap\overline{AI}$ (blue curve) and a second one for comparing the rankings of the disciplines that are citing them over time (red curve).

This figure shows that the disciplines cited by $\mathcal{P}\cap AI$ remain quite different from those cited by $\mathcal{P}\cap \overline{AI}$.
However the almost linear growth over time of the similarity of references is a signal of bibliographic homogenisation.
We observe also on this figure a similar behavior for the second similarity measure related to the disciplines citing the two corpora, which could be divided at first glance in two stages.
First the similarity stabilized at a very low value between 1970 and 1987, ie. the two corpora were cited by two very different sets of disciplines, then it has been steadily increasing since 1988, ie. the sets of disciplines that are showing interest for respectively the AI-related literature and the non-AI one, have become more intertwined over time.

\subsection{Neuroscience and AI: two different disciplinary landscapes}
\label{sec:results_disc_carto}

We now study the difference between the disciplinary environments of the AI-related corpus ($\mathcal{P}\cap AI$) and the non-AI one ($\mathcal{P}\cap\overline{AI}$) in our neuroscience dataset.
In particular, by considering the individual disciplines that are commonly cited by (or are citing) both of the two corpora, we compare to what extent these disciplines are actually cited by (are citing) each of them.
The goal of such an analysis is to identify the disciplines from which each corpus prefers to find information, ie. their references, and those into which they spread their knowledge, ie. their citations.

By applying the framework described in Section~\ref{sec:build_disciplinary_landscape}, we associate with each represented discipline $d$ in the references or citations of the corpora $\mathcal{P}\cap \overline{AI}$ and $\mathcal{P}\cap AI$, their time-aggregated ranks over decades, $\left(\tilde{r}^d_{\mathcal{P}\cap \overline{AI}}(T), \tilde{r}^d_{\mathcal{P}\cap AI}(T)\right)$, with $T=\left[t_0,t_0+10\right)$ a given time period where $t_0\in\lbrace 1970,1980,1990,2000,2010 \rbrace$.
These ranks are built with the total numbers of references to the discipline $d$ made by the corpora (references), or with the total numbers of citations given by $d$ to the corpora (citations, or impact) during the time period $T$.

\begin{figure}
    \centering
    \includegraphics[width=\textwidth]{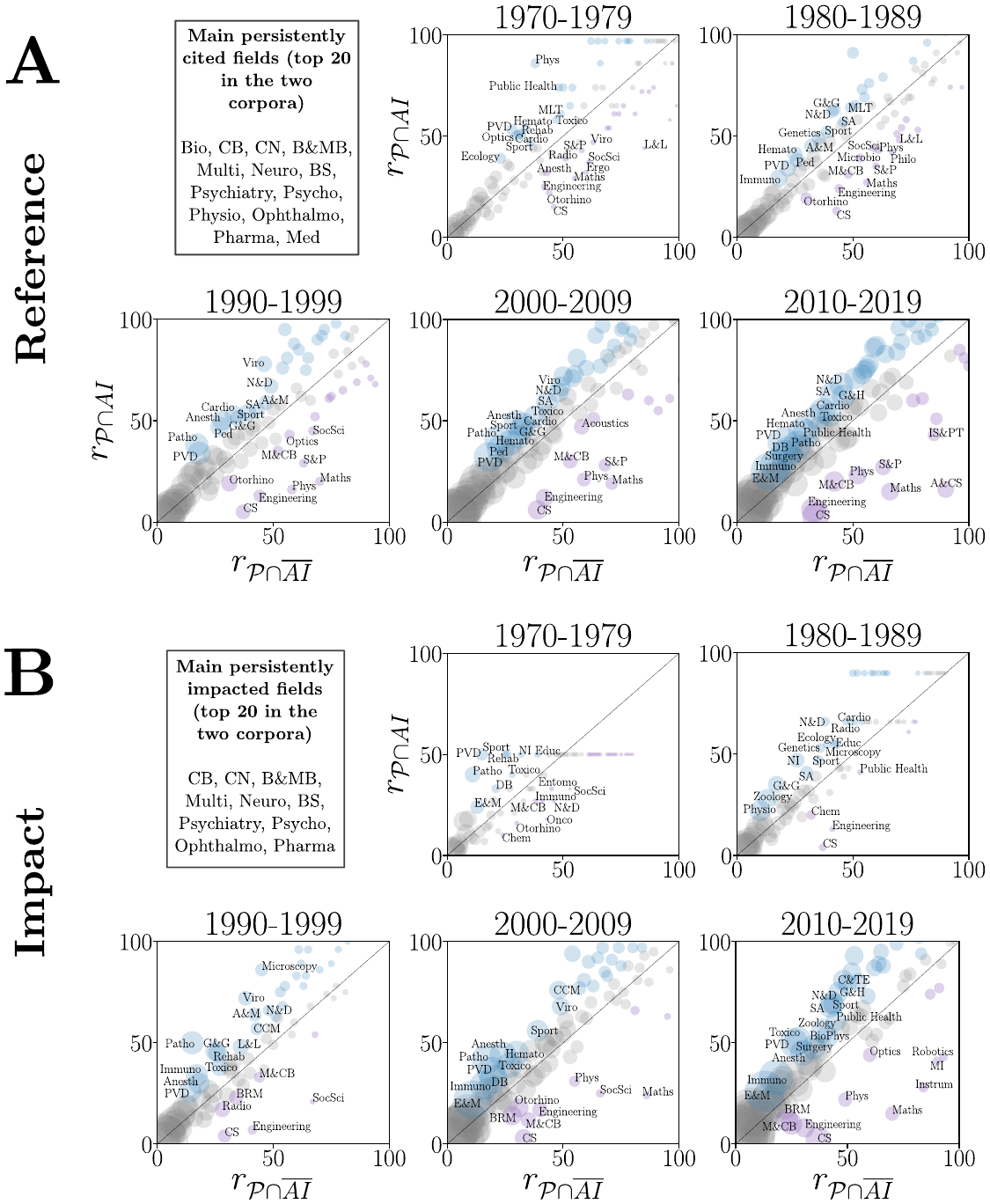}
    \caption{Time-aggregated ranking maps of the fields of study involved in publications \textit{cited by} both of the studied corpora $\mathcal{P}\cap AI$ and $\mathcal{P}\cap \overline{AI}$ (\textbf{A}), and in publications that are \textit{citing} them (\textbf{B}).
    The dashed lines show the diagonal where the rankings are exactly the same in the two corpora.
    Only are shown the most significant disciplines, with the condition that they commonly appear in the two corpora.
    The sizes of the discs, based on their empirical number of citations in the two different corpora, are normalized to compare their respective citation weights within each corpus.
    Grey points correspond to disciplines confined in the area between the two lines of respective equations $r_{\mathcal{P}\cap AI} = r_{\mathcal{P}\cap\overline{AI}}+\tau$ (above the diagonal) and $r_{\mathcal{P}\cap AI} = r_{\mathcal{P}\cap\overline{AI}}-\tau$ (below the diagonal), with $\tau=10$, and with no important variations of positions from one period to one another.
    The most persistent ones over the 5 represented decades, with a rank lower than 20, are mentioned in the upper left box of each figure \textbf{A} and \textbf{B}.
    Finally, blue points are the most preferred disciplines of the $\mathcal{P}\cap \overline{AI}$ corpus while purple ones are those for the corpus $\mathcal{P}\cap AI$.
    Abbreviations of the apparent disciplines are given in App.~\ref{app:wos_abb_table}.}
    \label{fig:rankings-2D_plots}
\end{figure}

We then locate the cited or citing disciplines in the 2D space of rankings associated with each corpus respectively, as shown in Fig.~\ref{fig:rankings-2D_plots}.
In one of the maps drawn on this figure, one colored disc represents a specific discipline $d$ that is located with its respective ranks in each corpus during the considered time period, in references (Fig.~\ref{fig:rankings-2D_plots}A) or citations (Fig.~\ref{fig:rankings-2D_plots}B).
Lower the value of one axis, better the associated rank, and then higher the number of collected citations along this axis.
From these maps we define the common interest area between $\mathcal{P}\cap AI$ and $\mathcal{P}\cap \overline{AI}$ as the region close to the diagonal where the two rankings of the fields of study are almost the same and not significantly varying over time.
These fields are colored in grey in these maps.
The fields that are more dispersed around the diagonal are represented with two different colors, the blue ones having a better rank in $\mathcal{P}\cap \overline{AI}$ than in $\mathcal{P}\cap AI$ and the purple ones having a better rank in $\mathcal{P}\cap AI$ than in $\mathcal{P}\cap \overline{AI}$.

\begin{figure}[htb]
    \centering
    \includegraphics[width=\textwidth]{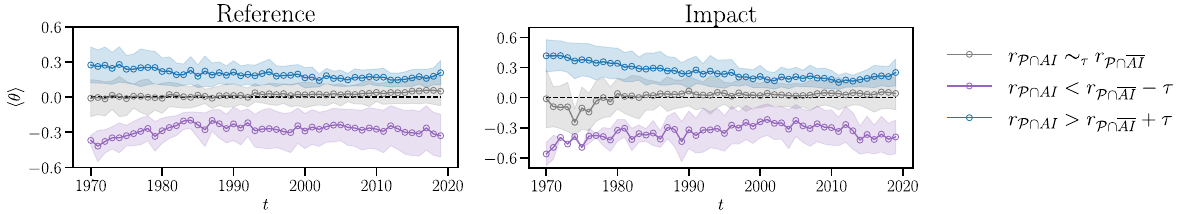}
    \caption{Temporal average angles produced by the disciplines in each area of reference's (left) and impact's (right) diagrams of Fig.~\ref{fig:rankings-2D_plots}, with respect to the diagonals represented here by a dashed line at $\langle\theta\rangle=0$.
    The colored area covering the curves are representing their respective standard deviation from the mean angle.
    The angles are expressed in radians.}
    \label{fig:rankings_angles}
\end{figure}

To characterize each of these three areas around the diagonal in the references (citations) of each corpus, we use the annual couples of ranks $r^d_R(t)$ ($r^d_I(t)$) of each represented discipline $d$ in $\mathcal{P}\cap \overline{AI}$ and $\mathcal{P}\cap AI$, and we compute the temporal angle $\theta(t)$ that indicates its temporal deviation from the diagonal, here corresponding to the null angle $\theta=0$.
Then we compute inside each zone and at each year the average angle $\langle\theta\rangle$ and the standard deviation from the latter, both represented in Fig.~\ref{fig:rankings_angles}.

The disciplinary composition of the common interest area, concentrated around the diagonals of all the maps, includes rather the same fields at each decade, both on references' and impact's sides.
According to the common ones that are more persistent over decades inside the references and received citations (see upper left boxes in Figs.~\ref{fig:rankings-2D_plots}A and \ref{fig:rankings-2D_plots}B), the core of the observed citation dynamics lies in neuroscience and is composed by disciplines that are mainly associated with medicine and biomedical research, such that \textit{Biochemistry \& Molecular Biology}, \textit{Behavioral Sciences}, \textit{Clinical Neurology}, \textit{Physiology}, \textit{Cell Biology}, \textit{Psychology}, \textit{Psychiatry} and \textit{Ophthalmology}.
This zone is thus coinciding with the definition of neuroscience given by neuroscientists themselves, namely it ``include[s] all fields that are involved with the study of the brain, the behaviors that it generates, and the mechanisms by which it does so, including cognitive neuroscience, systems neuroscience and psychology'' (\cite{hassabis_neuroscience-inspired_2017}, p.~245).
This zone is also accompanied by disciplines whose ranks are more variable and that are associated to more technological aspects of neuroscience, such as \textit{Computer Science}, \textit{Engineering}, \textit{Radiology}, \textit{Neuroimaging} and \textit{Audiology \& Speech-Language pathology}.
A description of the evolution of the disciplinary landscape inside this zone is detailed in App.~\ref{app:disc_comp_cia}.

The special disciplinary ecosystem of the non-AI corpus $\mathcal{P}\cap\overline{AI}$, in blue in the rankings maps of Fig.~\ref{fig:rankings-2D_plots}, is also centered around biomedical fields of study that tend to be close to the disciplines that characterize the whole neuroscience as mentioned above.
In particular, as shown in Fig.~\ref{fig:rankings_angles}, the mean angles and the standard deviations of the disciplines in the associated zone are asserting a global concentration of them toward the common interest area, which includes on one hand the most influential fields of research (reference) and on the other hand the core of those that show the same interest for the two studied corpora (impact).
We also notice in the reference's angle plot of Fig.~\ref{fig:rankings_angles} a deviation from the diagonal of the disciplines in this zone of references since 2010, that is also shared by the disciplines in the common interest area of references.
This suggests a recent shift of references shared by the two corpora toward fields of study preferred by the non-AI corpus.

Conversely, the special disciplinary ecosystem of $\mathcal{P}\cap AI$ evolves differently by representing the mathematical, computational and technological part of neuroscience since 1970.
The regular references to \textit{Computer Science}, \textit{Physics}, \textit{Statistics \& Probability}, \textit{Mathematical \& Computational Biology} and \textit{Engineering} show a large influence of technological-oriented research in this particular AI research in neuroscience.
Progressively concentrated toward the common interest area between 1970 and the late 1980 (see Fig.~\ref{fig:rankings_angles}), these references preferred by $\mathcal{P}\cap AI$ become more further away from the common interest area and more dispersed after this period, thus indicating a cognitive differentiation of references on which the AI-related corpus $\mathcal{P}\cap AI$ is drawing upon from the non-AI corpus $\mathcal{P}\cap \overline{AI}$.
In addition, while neuroscience and associated medical fields -- as \textit{Clinical Neurology} and \textit{Neuroimaging} -- remain the primary stakeholders in the AI research conducted within it, the latter appears to be of varied interest, since the 1980s, to a subset of disciplines which do not place as much emphasis on works in the non-AI corpus and which are common to those cited preferentially by the AI-corpus over time, such as \textit{Computer Science}, \textit{Engineering} and \textit{Mathematics for Computational Biology}, the last one exhibiting especially a spectacular increase of its own rank between 1990 and 2019 (see Figs.~\ref{fig:rankings-2D_plots}A and B\footnote{Fig.~\ref{fig:rankings-2D_plots}B shows also a particular proximity of the AI-related corpus with the field of chemistry, here represented with the JSCs \textit{Chemistry} and \textit{Biochemical Research Methods}, the latter giving more citations than the former since 1990.}).
Moreover, these special disciplines that cite most the corpus $\mathcal{P}\cap AI$ in Fig.~\ref{fig:rankings-2D_plots}B are as dispersed as the most cited ones by this corpus in Fig.~\ref{fig:rankings-2D_plots}A, and remain over time less close to the common interest area than the disciplines preferring the other corpus $\mathcal{P}\cap \overline{AI}$ (see the corresponding annual evolution of the average angle $\langle\theta\rangle$, colored in purple in the impact side of Fig.~\ref{fig:rankings_angles}).
All these results thus show that AI research in neuroscience is situated in a particular disciplinary environment that is not shared by the core of the neuroscience field.

\begin{figure}[ht]
    \centering
    \includegraphics[width=0.9\textwidth]{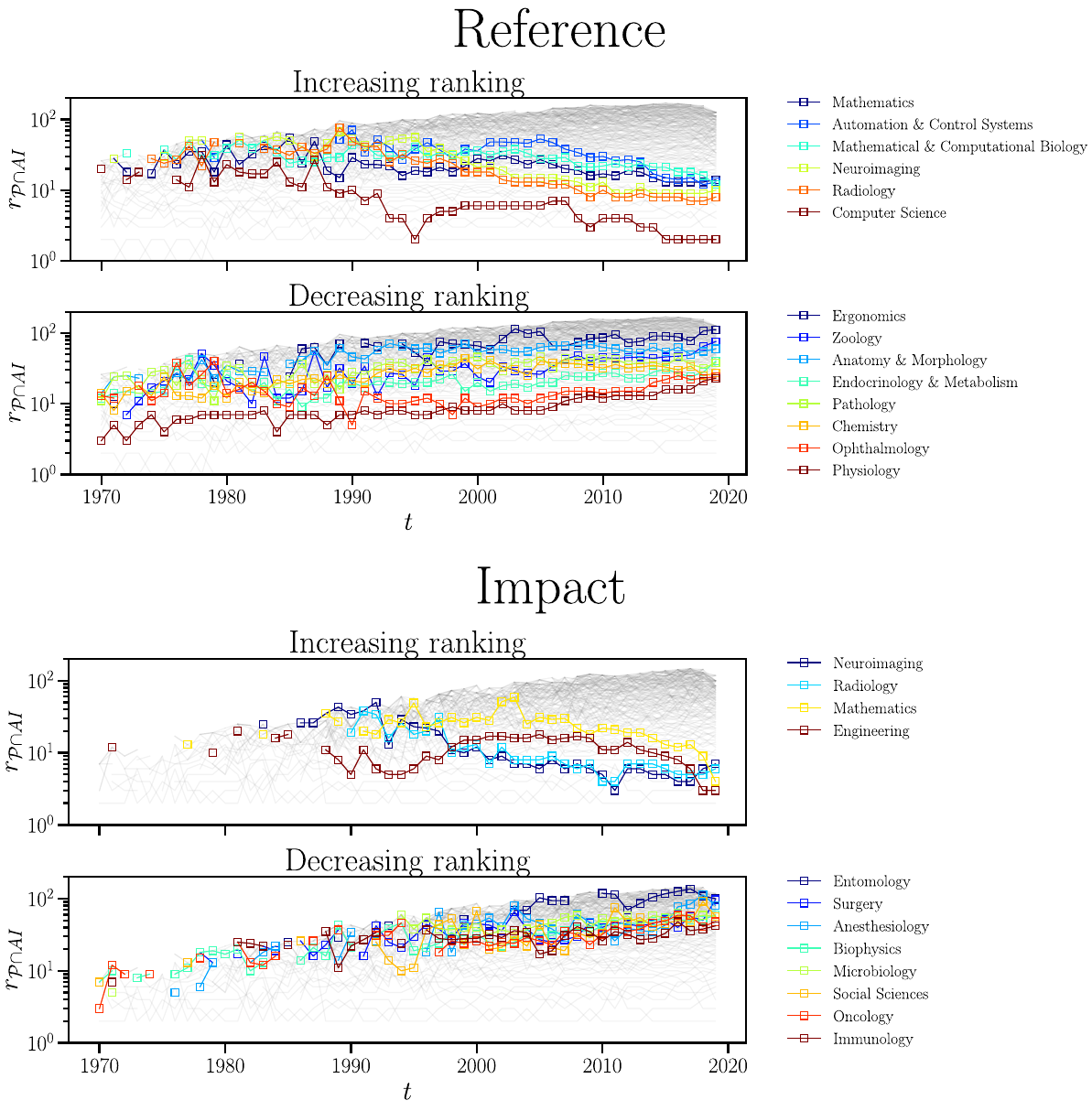}
    \caption{Time evolution of the ranking $r_{\mathcal{P}\cap AI}$ of the disciplines that are mostly represented by the corpus $\mathcal{P}\cap AI$ in the references (top) and citations (bottom), ie. situated in the lower zone of the maps shown in Fig.~\ref{fig:rankings-2D_plots}.
    Only the curves with most significant evolution are highlighted with colors.}
    \label{fig:special_ranking_AI}
\end{figure}

By regarding the disciplines with the most significant rank evolution within the AI ecosystem $\mathcal{P}\cap AI$, Fig.~\ref{fig:special_ranking_AI} shows the rise of \textit{Neuroimaging} and \textit{Radiology} both on references' and impact's sides, as already observed in the common interest area, and also the progressive domination of \textit{Computer Science}.
This figure shows also the decrease of the influence of fields of research linked to the disciplinary orientations of neuroscience and especially of the corpus $\mathcal{P}\cap \overline{AI}$.
This is also a sign of a progressive differentiation of the research supported through AI from the rest of neuroscience.
The case of \textit{Physiology}, which contributed to the foundations of neuroscience \citep{cooper_cognitive_2010}, is particularly eloquent with its distancing in time from the highest positions in the reference ranking of the corpus $\mathcal{P}\cap AI$ since the 2000s.
This observed retreat of master neuroscience disciplines, that had a strong influence in the building of AI in this field since the 1970s, suggests that a social transformation occurred in this special research during the studied temporal period.
Perconti and Plebe (\citeyear{perconti_deep_2020}) mention such a transformation, in which AI in neuroscience was a matter for biomedical specialists before becoming an object of study and technological developments for engineers.
This will be shown empirically in the following sections.

\subsection{An AI literature confined in a small set of journals}
\label{sec:results_journals}

Another factor of differentiation of the AI research from the core of neuroscience lies on the set of journals in which the former are mainly published.
Here we analyze how AI is promoted within the 421 journals included in the whole neuroscience dataset, and who are the authors heading to the neuroscience journals publishing most AI research.

\begin{figure}[htb!]
    \centering
    \includegraphics[width=1\textwidth]{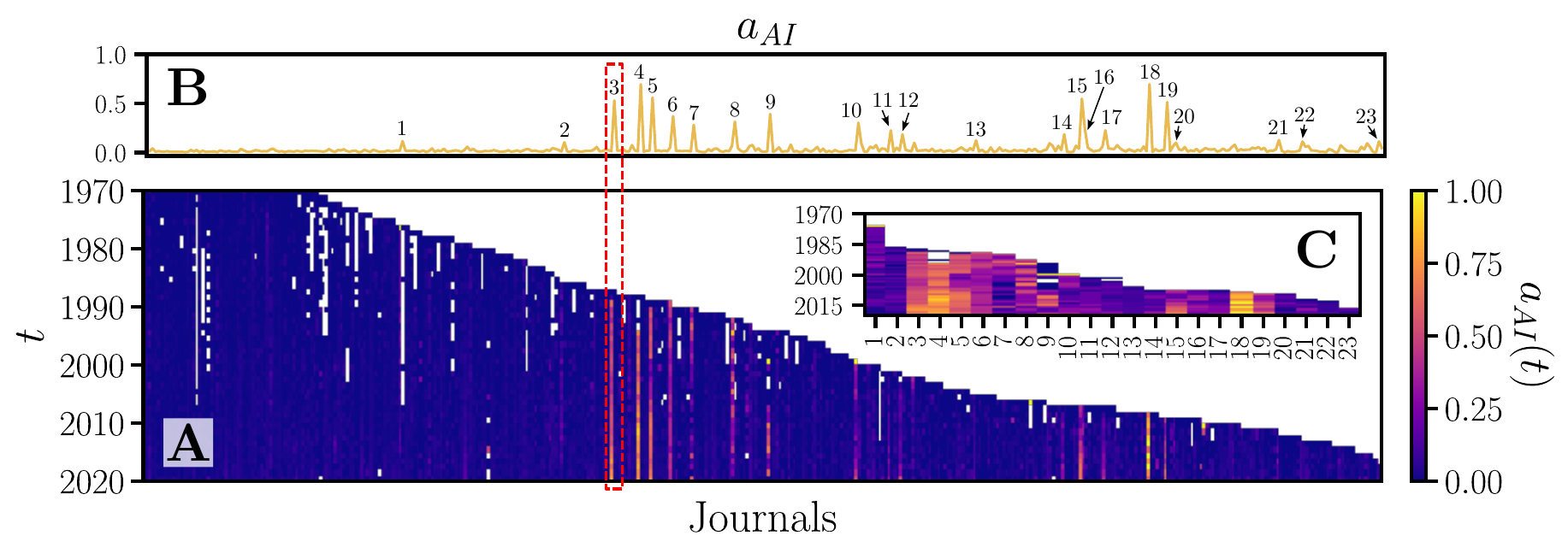}
    \begin{tabular}{l p{0.5\linewidth}p{0.2\linewidth}rr}
        \hline
         & Journal & Launch year in the database & $a_{AI}$ \\
        \hline
        1 & Cognitive Science & 1976 & 0.12 \\
        2 & Journal of Sensory Studies & 1986 & 0.10 \\
        3 & Neural Networks & 1987 & 0.53 \\
        4 & Neural Processing Letters & 1988 & 0.70 \\
        5 & Neurocomputing & 1989 & 0.56 \\
        6 & Neural Computation & 1989 & 0.37 \\
        7 & Network: Computation in Neural Systems & 1990 & 0.28 \\
        8 & Adaptive Behavior & 1992 & 0.31 \\
        9 & Neural Network World & 1994 & 0.39 \\
        10 & Cognitive Systems Research & 1999 & 0.30 \\
        11 & IEEE Transactions on Neural Systems and Rehabilitation Engineering & 2001 & 0.22 \\
        12 & Neuroinformatics & 2001 & 0.19 \\
        13 & Plos Computational Biology & 2005 & 0.12 \\
        14 & Frontiers in Neuroinformatics & 2007 & 0.19 \\
        15 & Computational Intelligence and Neuroscience & 2007 & 0.55 \\
        16 & Cognitive Neurodynamics & 2007 & 0.24 \\
        17 & Frontiers in Computational Neuroscience & 2007 & 0.23 \\
        18 & Evolutionary Intelligence & 2008 & 0.70 \\
        19 & Cognitive Computation & 2009 & 0.51 \\
        20 & Topics in Cognitive Science & 2009 & 0.10 \\
        21 & Journal of Mathematical Neuroscience & 2011 & 0.13 \\
        22 & NeuroImage: Clinical & 2012 & 0.11 \\
        23 & Biological Psychiatry: Cognitive Neuroscience and Neuroimaging & 2016 & 0.11 \\
        \hline
    \end{tabular}
    \caption{Top diagrams: \textbf{A}: Temporal AI activity of the journals. One square at the position $(j,t)$ of the diagram is the share of AI publications in the journal $j$ at time $t$. One column of the plot is the AI activity of one journal. 
    \textbf{B}: Distribution of the global AI activity of the journals over all their publications up to 2019. The horizontal axis is corresponding to the journals axis of the bottom plot A, as shown by the red dashed zone.
    \textbf{C}: Zoomed activities of the 23 most active journals in AI research, with a global AI-activity higher than 10\%. They are indicated in the bottom table.}
    \label{fig:journal_activity}
\end{figure}

For each journal, based on its respective entire publication history, we compute the temporal share $a_{AI}(t)$ of AI-related publications (called also \textit{temporal AI activity}), and its global share $a_{AI}$ of such publications since its launch year (called also just \textit{AI activity}).
These scores are represented in Fig.~\ref{fig:journal_activity}.
We then identify the journals with $a_{AI}$ higher than 10\%, whose names and launch years are given in the table in the bottom of this figure.

According to the evolution of the AI activities of neuroscience journals, the AI-related research in this field is concentrated around a small subset of journals providing development in computational techniques mainly linked to neural networks and cognition.
These journals are quite representative of the connectionist wave of AI which is active in neuroscience since the late 1980s \citep{cardon_neurons_2018}.
They are representing 32,7\% of the scientific production of AI in neuroscience during the whole period 1970-2019, according to our dataset, the rest being distributed among the 398 other journals with a much smaller AI-activity (less than 10\%).

The spectrum of $a_{AI}$ (top yellow curve in Fig.~\ref{fig:journal_activity}) shows that the launches of the most AI active journals in the neuroscience field are concentrated around three periods.
The first one spans the period 1987-1994 with especially 7 journals (no. 3 to 9 in the table of Fig.~\ref{fig:journal_activity}) whose $a_{AI}$ is higher than 28\% and whose scopes are oriented toward computational neuroscience and the use of neural network formalism for complex calculations, especially the simulation of cognition on neural systems.
These journals, except no. 7 and 9 (respectively \textit{Network: Computation in Neural Systems} and \textit{Neural Network World}) are showing well-sustained activity on AI research.
This period falls commonly into the second ``AI winter'', when AI-research funding and the production of scientific results and associated technological solutions were at their lowest for a second time \citep{cardon_neurons_2018,schuchmann_history_2019}.
Paradoxically, neuroscience are especially active in such research in this period, as shown by the strong, long-lasting AI-activity of these journals created in this period.

The second period includes the journals 10, 11 and 12, which were launched between 1999 and 2001.
The journal \textit{IEEE Transactions on Neural Systems and Rehabilitation Engineering} is especially oriented toward the development of computational methods and technological tools to capture the neural activity of the brain. 
This period is followed by another until 2007 which did not lead to the launch of AI-active journals.

Finally, the third period following the latter, between 2007 and 2009, saw the launch of numerous journals (no. 14 to 20) in a particularly short time, that are fostering research at the crossroads of neuroscience and cognitive science.
It includes especially the most AI active journal of our dataset, \textit{Evolutionary Intelligence}, whose scope is oriented toward evolutionary computation, a subset of the field of optimization.
Its temporal activity is also strongly sustained over the whole time period covered in our datasets.

\begin{figure}[htb]
    \centering
    \includegraphics[width=0.6\textwidth]{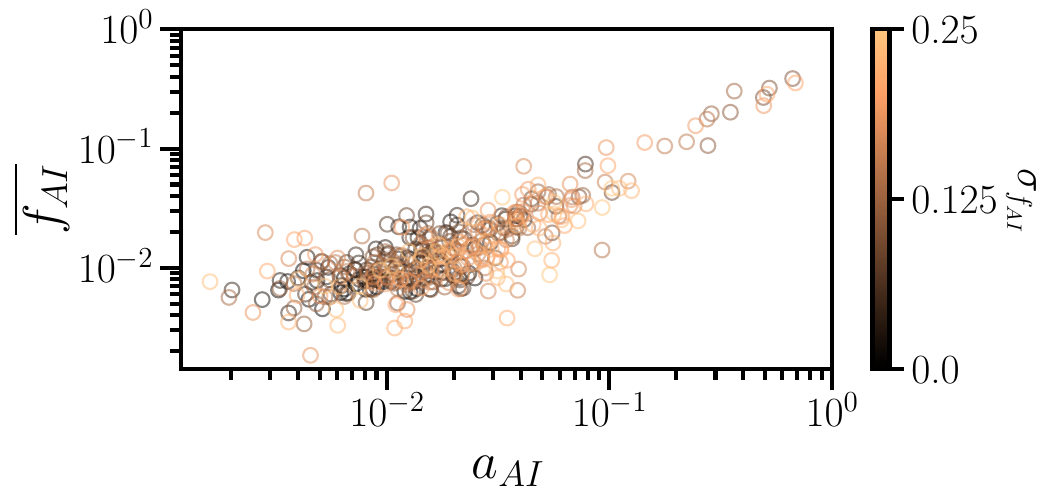}
    \caption{Correlation plot between the global AI activity $a_{AI}$ of the journals appearing in $\mathcal{P}$ and the average AI scores $\overline{f_{AI}}$ of the authors who are published in it between 1970 and 2019. 
    Each point represents one journal in the horizontal $a_{AI}$ space, colored by the standard deviation of the $f_{AI}$ scores of their associated authors. 
    The maximum average value of $\overline{f_{AI}}$ is situated around 0,4.}
    \label{fig:corr_authors_journals}
\end{figure}

Aggregating over time, Fig.~\ref{fig:corr_authors_journals} shows a linear correlation between the mean AI-activity in the journals and the average of the AI scores $f_{AI}$ of the authors that have published at least one paper in these journals.
The tail at highest $a_{AI}$ (higher than 0,1) is corresponding to the top 15 of the most AI-active journals given in Fig.~\ref{fig:journal_activity}.
This particular result thus unveils attraction of the authors publishing the most AI-related works of our dataset in the journals with a high AI-activity in time.
This suggests that AI research in neuroscience is done by a specialized scientific community inside this field with its own journals for communicating results.

\subsection{Who are the scientists making AI in neuroscience?}
\label{sec:results_authors}

\subsubsection{Disciplinary profiles in each quartile of the $f_{AI}$ distribution}

In this part we focus on the profiles of the scientists doing AI in neuroscience (in Fig.~\ref{fig:AIscore_distrib}, included in one of the quartiles $Q_0$, $Q_1$ or $Q_2$) and how they are inserted into the global authorship landscape of neuroscience.
In particular, we compare these AI practitioners with the other scientists in neuroscience who never published AI-related papers (included in the quartile $\overline{Q}$) under two aspects, namely (1) their \textit{disciplinary background} and (2) their \textit{disciplinary trajectory} in academia. 

We define the disciplinary background of one author as the set of unique disciplines corresponding to the journals in which he/she has published in his/her first year of academic life, namely the year of his/her very first publication(s).
For example, a fictitious author publishing his/her three first papers in the same year $y_0$ in two different journals labeled with disciplines $\{d_1\}$, $\{d_1,d_2\}$ and $\{d_3\}$ respectively, would have a disciplinary background composed of disciplines $\{d_1,d_2,d_3\}$.
In this way, another fictitious author who has published only one paper in his/her first year of academic life (actually the most frequent situation, occurring 64\% of times in the dataset) would have a disciplinary background composed only of the discipline(s) labeling the corresponding journal.

\begin{figure}[htb!]
    \centering
    \includegraphics[width=0.9\textwidth]{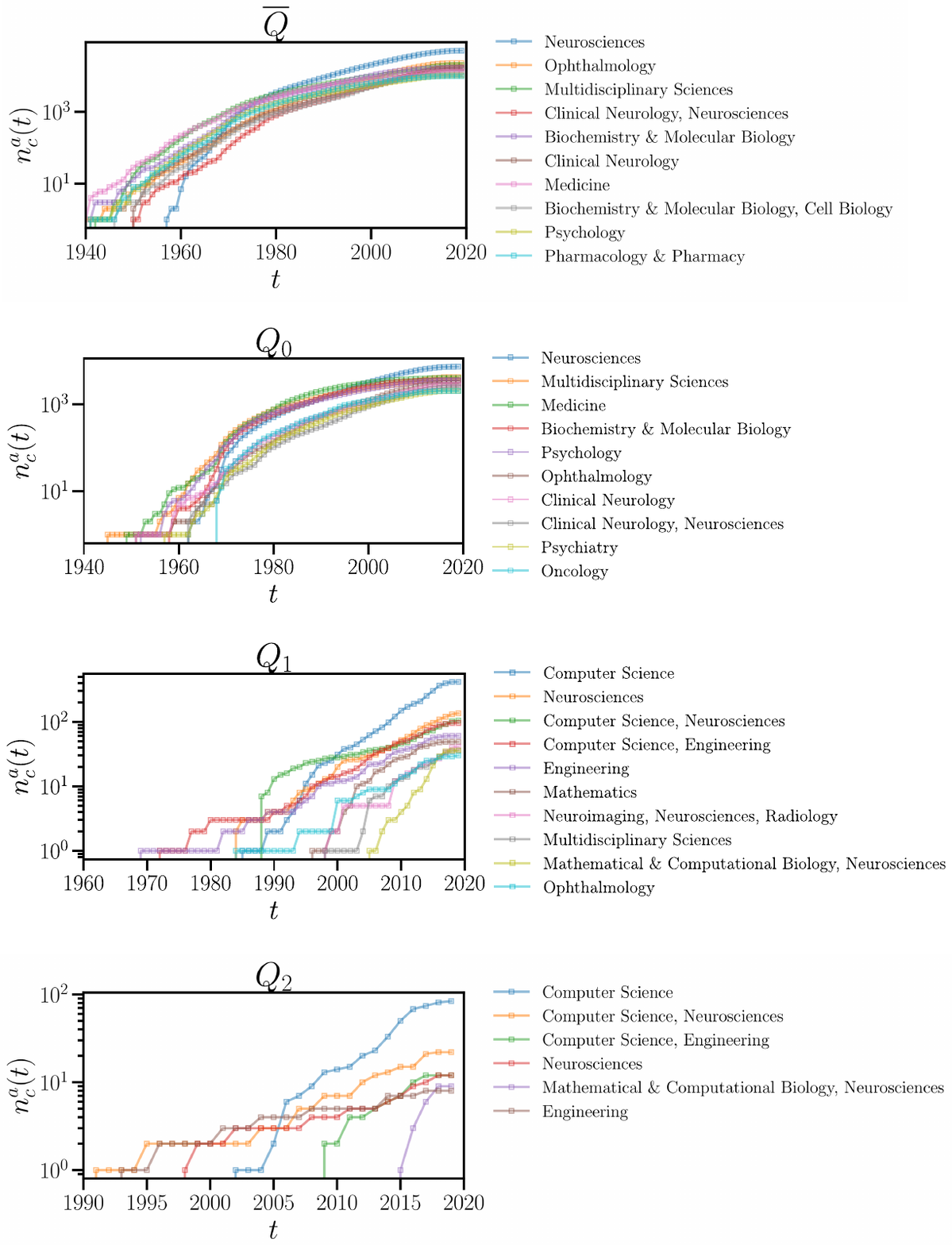}
    \caption{Cumulative number of new authors per disciplinary background for each quartile. For one plot, the point of the curve of the specialty $d$ at year $t$ is the number of authors who have published their first articles inside $d$ during their very first year of academic career, since the year of first appearance of $d$. Only the top 10 native specialties in 2019 are shown in $\overline{Q}$, $Q_0$ and $Q_1$, and the top 6 for $Q_2$ because of the insignificance of the following ones.}
\label{fig:disciplinary_origins}
\end{figure}

Then we compute for each quartile the temporal cumulative number of new scientists trained in each represented disciplinary background in this quartile, as shown in Fig.~\ref{fig:disciplinary_origins}.
In this way we assess for each quartile the main native specialties in which authors have first published.

Fig.~\ref{fig:disciplinary_origins} shows that the profiles included in $Q_0$ and $\overline{Q}$ overall the period 1940-2019 are very similar. 
They are mainly confined in biomedical research around neuroscience, as well as in \textit{Multidisciplinary Sciences} which is represented by 91 international journals with a broad topical diversity.
We recognize also the main fields of research that are shaping the common interest area of the disciplinary ecosystems of the two separated AI and non-AI corpora (see the most persistent disciplines in references and citations in Fig.~\ref{fig:rankings-2D_plots}).
These two profiles best represent the disciplinary spectrum of neuroscience itself.
We notice in addition the spectacular increase of the \textit{Neurosciences} curve in the two plots until the 1970s (after having emerged in 1957 for $\overline{Q}$ and in 1962 for $Q_0$), followed by a quasi-linear progression until 2015.
This boom of neuroscience profiles in these quartiles suggests that modern neuroscience is progressively being institutionalized as a well-structured discipline in science.

The profiles of $Q_1$ and $Q_2$ shown in Fig.~\ref{fig:disciplinary_origins} are at the opposite of the previous ones, coming at most from fields of research related to \textit{Computer Science}.
The specialty \textit{Computer Science, Neuroscience}, which emerged in 1988 for $Q_1$ and in 1991 for $Q_2$ indicates the rising of a group of scientists that are specialized into computation in neuroscience.
Notice that the behavior of the \textit{Neurosciences} curve within $Q_1$ in Fig.~\ref{fig:disciplinary_origins} appears later than computational and engineering profiles, suggesting also that AI-related knowledge and technological tools penetrate progressively the global neuroscience field.

\begin{figure}[htb!]
    \centering
    \includegraphics[width=0.85\textwidth]{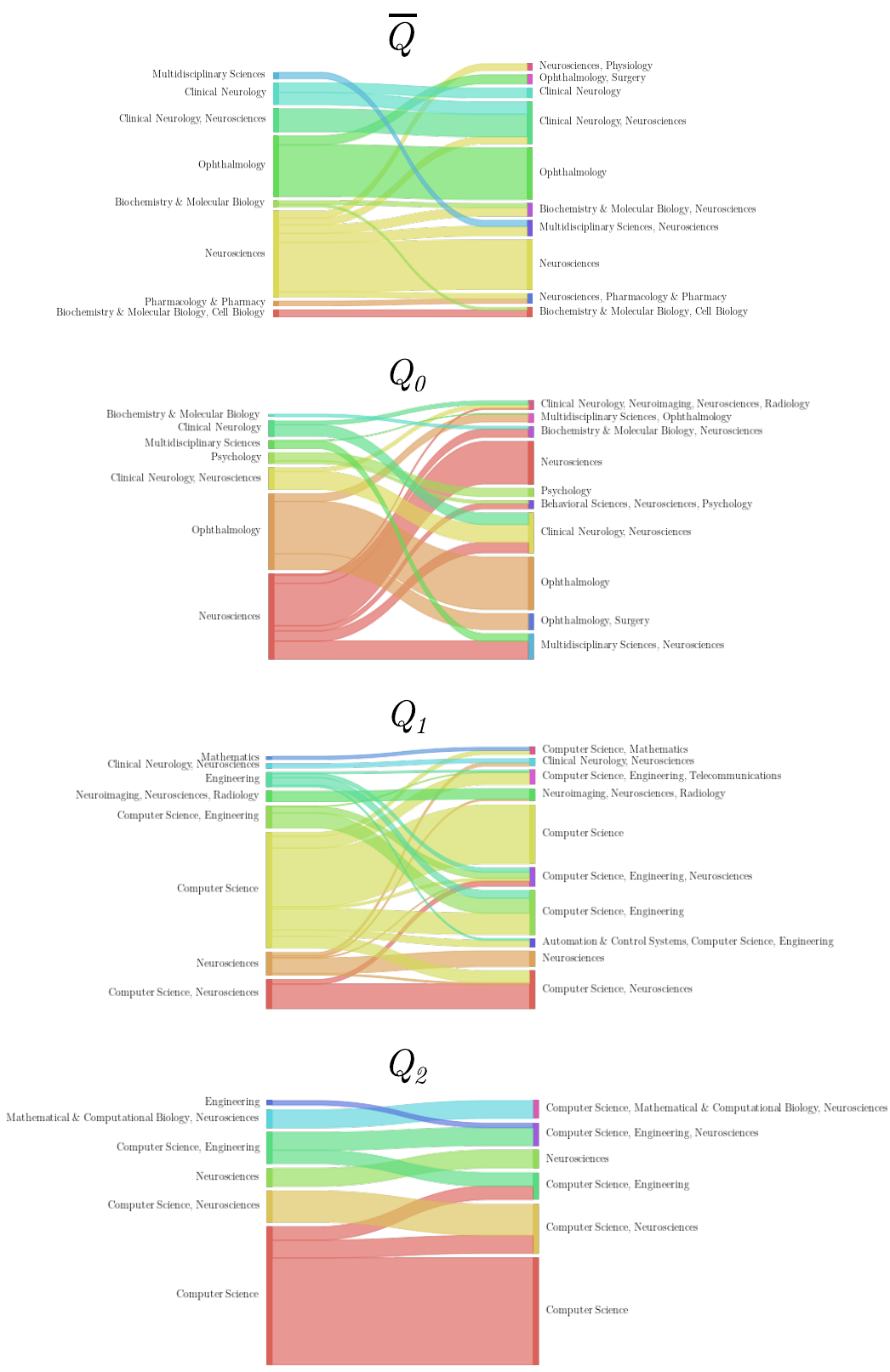}
    \caption{Significant disciplinary trajectories of the authors belonging in each quartile. On each plot lie on the left the most significant disciplinary backgrounds, and on the right lie the career-related disciplines up to 2019.}
\label{fig:disciplinary_trajectories}
\end{figure}

From the subset of authors belonging to the most frequent disciplinary background in each quartile, we consider their respective disciplinary profile as the disciplines corresponding to the journals in which they publish throughout their scientific life -- that would not be ended for the youngest still publishing in 2019.
We therefore draw from these two parameters -- disciplinary background and career-related disciplines -- the typical disciplinary trajectories in each quartile, which are shown in Fig.~\ref{fig:disciplinary_trajectories}.
Since the backgrounds are built from the publications at a given year, the authors who began their career in 2019 would have a disciplinary profile corresponding to it.
For avoiding an over-representation of some confined disciplinary trajectories due to these newcomers, we therefore consider only the authors who began at the latest in 2018.
For the sake of clarity we select only the most significant trajectories.

As a confirmation of the disciplinary background shown in Fig.~\ref{fig:disciplinary_origins}, Fig.~\ref{fig:disciplinary_trajectories} shows that the studied authors in $\overline{Q}$ and $Q_0$ are involved in a similar disciplinary landscape centered around \textit{Ophthalmology}, \textit{Clinical Neurology} and general \textit{Neurosciences}, while $Q_1$ and $Q_2$ are more confined into \textit{Computer Science} and \textit{Engineering}.
In particular, by considering neuroscience as the subset of disciplines composed of the WOS disciplines \textit{Neurosciences}, \textit{Clinical Neurology} and \textit{Neuroimaging}, we observe that $\overline{Q}$ and $Q_0$ are more involved in that field of research, with respectively 77\% of the authors in the first and 78\% of those in the second with a disciplinary profile that includes one or more of the fields of research associated with neuroscience.
On the contrary, 45\% of the authors in $Q_1$ and 42\% of the authors in $Q_2$ have a disciplinary profile that includes such JSCs.
These two last quartiles are therefore mainly detached of the neuroscience goals given the disciplinary backgrounds and profiles of their main respective authors.
These results thus show that neuroscience is a field that brings together heterogeneous profiles who seem to serve different epistemic objectives inside and outside neuroscience \citep{sedooka_paradoxe_2015}.

Nonetheless we notice two special things about these last quartiles.
First the authors who began into the specialties including one or more neuroscience-related disciplines tends to continue in the same field of research, which is sometimes interdisciplinary, such as the combinations \textit{Computer Science, Neurosciences} and \textit{Neuroimaging, Neurosciences, Radiology}.
This means that these scientists are trained into a disciplinary context centered around neuroscience. 
Second, some authors who began into \textit{Computer Science}, \textit{Engineering} or mathematics-related disciplines -- who seem to be more distant from neuroscience in their disciplinary backgrounds -- continue into neuroscience, as shown by the combination of their originating field(s) of research with neuroscience ones in their disciplinary profiles in Fig.~\ref{fig:disciplinary_trajectories}, for instance \textit{Computer Science, Engineering} leading to \textit{Computer Science, Engineering, Neurosciences}.
The late emergence of interdisciplinary profiles in computer science and neuroscience, who are more involved in neuroscience in general, such as those in $Q_2$, also testify that the AI community is taking root in the global neuroscience landscape, the more recent profiles in neuroscience becoming \textit{insiders} in this new technological specialty.
All these scientists thus represents a special labor force for neuroscience whose main expertise lies in AI, and more generally in mathematical, computational and technological tools \citep{perconti_deep_2020}.

\subsubsection{Structure of the collaboration network between quartiles}
\label{sec:results_authors_collab}

We investigate in the following how AI practitioners are distributed within the neuroscience community, and how are shaped the collaborations between them, especially between the different kinds of scientific profiles described before.

\begin{figure}[htb!]
    \centering
    \begin{tabular}{ll}
    \includegraphics[width=0.9\textwidth]{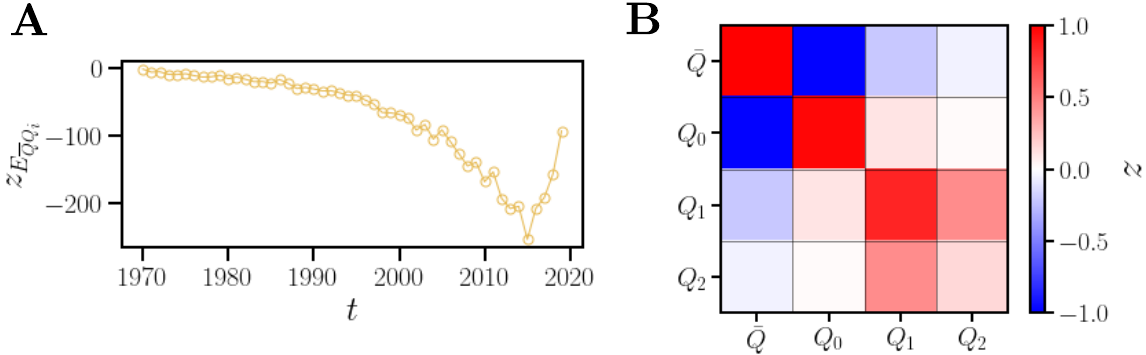}
    \end{tabular}
    \caption{\textbf{A}: Temporal $z$-score applied to the edges in TCN between the scientists in $\overline{Q}$ and those in the other $Q_i$, here aggregated together under the notation $Q_i$.
    \textbf{B}: $Z$-score matrix of the edges between each quartile in the time-aggregated collaboration network ACN. The values are normalized with the absolute maximum one in the matrix.}
\label{fig:collabNet_quartile}
\end{figure}

We first consider the temporal collaboration network TCN built in Section \ref{sec:build_collaboration_network}, and we evaluate the temporal standardized share of edges between scientists belonging respectively to $\overline{Q}$ and to all other quartiles $Q_i$ (see Section~\ref{sec:z-score_collab} for the calculation of such score).
This score is showed in Fig.~\ref{fig:collabNet_quartile}A, from which we deduce two facts.

First, all its values are lower than 0 over time, meaning that the number of edges between the two ensembles $\overline{Q}$ and $\{Q_0,Q_1,Q_2\}$, is lower than the average one computed from several random distributions of all the edges between the authors in TCN.
This fact thus shows that the AI practitioners within neuroscience and the other neuroscientists maintain few collaborations as TCN grows over time.
Second this situation worsens over time, with a steady decline until the 1990s followed by a steeper one toward 2015.
Although this tendency has reversed since 2015, these results indicates that the neuroscientists making AI, ie. in $Q_i$ and not in $\overline{Q}$, are shaping an almost independent community inside neuroscience by widening progressively a gap with other neuroscientists belonging to the set $\overline{Q}$.
Furthermore, concerning especially the authors in $Q_0$, who have similar disciplinary backgrounds and trajectories as the authors in $\overline{Q}$ (see Figs.~\ref{fig:disciplinary_origins} and \ref{fig:disciplinary_trajectories}), the results advanced above also suggest that the scientists in $Q_0$ became progressively \textit{outsiders} from the subset $\overline{Q}$ in the history of neuroscience by moving closer to AI -- even if the scientists in $Q_0$ have few AI-related publications.

This social divide is confirmed with Fig.~\ref{fig:collabNet_quartile}B, which represents the panorama of the links shared between the different quartiles in ACN (see also Section~\ref{sec:z-score_collab}).
This diagram represents a stabilized situation observed since the end of the 1990s (see Fig.~\ref{fig:collabNet_quartile_evol} in App.~\ref{app:collabNet_quartile_evol}), where the scientists in $Q_i$ cosign more with one another than with $\overline{Q}$ while the scientists in the second ensemble prefer to collaborate with one another as well.
This social separation around AI research does not produce, however, a strict knowledge divide in neuroscience, as demonstrated before with the temporal similarity index of disciplines impacted directly by neuroscience in Fig.~\ref{fig:jaccard}, as well as with the temporal evolution of the disciplinary landscapes in Fig.~\ref{fig:rankings-2D_plots}, both indicating meaning that the quartiles $Q_i$ export their knowledge in the whole neuroscience as much as the quartile $\overline{Q}$.

We also observe in Fig.~\ref{fig:collabNet_quartile}B a polarization within the subset of AI practitioners in neuroscience. 
Indeed, the authors belonging to $Q_1$ and $Q_2$ are more strongly connected together than with those in $Q_0$.
This can be explained by the disciplinary proximity of the authors in $Q_1$ and $Q_2$ observed in Fig.~\ref{fig:disciplinary_origins}, especially the prominence between the late 1980s and the early 2000s of full computer scientists and hybrid profile publishing in journals labeled as \textit{Computer Science, Neurosciences}.
Furthermore, the links between $Q_1$ and $Q_0$ are much more important than those between $Q_2$ and $Q_0$.
The scientists in $Q_1$ thus appears to be the most interdisciplinary by assuring the bridge between these differentiated profiles inside AI community.
These other \textit{outsiders} in neuroscience are especially driving the diffusion of AI in neuroscience from computation to medical and clinical applications, given their disciplinary trajectories shown in Fig.~\ref{fig:disciplinary_trajectories}.

\section{Discussion}
\label{sec:discussion}

In this article we have explored the penetration of AI in neuroscience through the co-development of the two fields and especially the construction of a technological specialty centered around AI inside neuroscience.
To do this we have conducted a scientometric analysis of an exhaustive bibliometric database that aims to represent at best neuroscience research between 1970 and 2019.
This analysis relies on many indicators. 

First, we have conducted a comparative analysis of the egocentric citation networks associated with respectively AI research and non-AI research within neuroscience.
We have shown that the sets of disciplines influencing respectively the AI-related works and the non-AI ones tend to be similar over time, as well as those impacted by these works.
AI thus becomes epistemically embedded in neuroscience over time, by reproducing citation patterns that are characterizing the field -- ie. toward the disciplines that structure the field --, and by increasingly impacting the disciplines that show the most interest in this field -- including neuroscience itself.

However, we have shown at micro-scale a progressive cognitive differentiation of the AI research from the non-AI one within neuroscience, based on both the employed bibliographic references in each of the two fields and their respective ensemble of impacted disciplines.
We especially have observed specialization of AI research toward computer science, mathematics and engineering, while the core of neuroscience research draws upon biomedical and clinical research fields related to it.
This differentiation is also produced by the neuroscience journals landscape, in which a small set of 23 journals, the most active in AI research in neuroscience, are representing those STEM fields in references and are gathering the scientists who are the most active in AI in neuroscience as well.
Furthermore, by observing the fields impacted by the two corpora under study, we show that AI research targets rather the same fields of research as those cited by it.
The epistemic integration of AI into neuroscience that was observed at first glance thus remains partial because of a specialty demarcation process of the former from the latter.

In a second step we explained this specialization by conducting a relational approach based on the scientific profiles of the authors of the studied database.
By distinguishing the AI practitioners in neuroscience from the other neuroscience specialists, we have shown that the first tend to not maintain links with the second in the temporal co-authorship network including the main collaborations in the field since 1970.
With the previous results we concluded that AI researchers are outsiders in neuroscience.
We especially have distinguished two classes in this group of scientists, namely a first wave of outsiders including authors who are trained in the main disciplines that are shaping the neuroscience field since the 1940s and who have a low AI activity, and a smallest second wave which emerged around the 1980s and that is including authors who are trained in other disciplines that are not represented in the former group, such as computer science and engineering, and who exhibit the highest activities in AI research in general (inside and outside neuroscience).
Furthermore this second group is not the most involved in the field of neuroscience, its members keeping to publish within their original disciplines that are mainly STEM ones.

This social polarization inside the AI practitioners in neuroscience suggests that AI becomes over the years a set of technologies that need to be shaped not only by neuroscientists themselves, but also with the help of scientists coming from outside neuroscience, or from within the discipline but with an interdisciplinary background, and who present specific expertise about AI itself.
Since we considered AI as a global research-technology in science \citep{shinn_transverse_2002, marcovich_science_2020,  hentschel_periodization_2015}, these results thus are a consequence of the diffusion of AI outside its originating STEM disciplines and throughout the science system \citep{gargiulo_meso-scale_2023}.
The second wave of AI researchers described before represents the mobility of such experts toward other fields of research, in order to propose and integrate the associated knowledge and technologies to achieve some of the disciplinary objectives of these receiving fields or, less ambitiously, to solve some technical problems that could not be solved with more conventional tools.

These results thus illustrate quite well the generic property of AI when applied in neuroscience, which is producing a social and cognitive differentiation inside the latter \citep{shinn_transverse_2002}.
However we could question the relevance of this criterion in this case, where only 3\% of publications in our dataset involves AI according to our keyword filter introduced in Section~\ref{sec:build_dataset}.
This leads especially to ask how spread is AI across the topics covered by neuroscience over time, ie. if it is present in all subfields of neuroscience or concentrated around a few ones.
With such a topic space, and by reusing data about authors, the co-signature network and the citation network, we could evaluate more precisely how much universal AI is inside neuroscience through the distributions of authors and citations in this topic space, and then deduce the propensity of AI to fit with the knowledge and methods associated with some topics rather than others.

Nonetheless, the dynamical process of integration of AI in neuroscience exhibits some differences regarding the global history of AI in science depicted in \citep{gargiulo_meso-scale_2023}.
Although the development of AI in science is statistically characterized by a disciplinary closure around STEM disciplines between 1980 and 2010, AI continued its interactions with neuroscience for its own epistemic purposes in this period, as shown by a large number of journals created in this period and that are very active in AI research (see Section~\ref{sec:results_journals}), and by its global citational impact in the discipline (see Sections~\ref{sec:results_disc_carto} and \ref{sec:results_disc_jaccard}).
However, a social closure was simultaneously occurring and accentuated inside the field, where not everyone is finally using AI at all, even in recent days.
According to Shinn and Joerges (\citeyear{shinn_transverse_2002}), this is typical of a differentiation process observed in the conception phase of an instrument.

The penetration of AI inside a single discipline thus could also be described as another underlying dynamical process of development of the associated knowledge and instruments following the four steps of Hentschel (\citeyear{hentschel_periodization_2015}), included inside the diffusion phase of the original instrument.
For instance, the AI developed inside neuroscience, potentially different from the originally what is produced in STEM disciplines, would be also adapted to furnish other capabilities in neuroscience first, but also in other disciplines or fields of research afterwards if they judge it useful for their own goals.
Under such an hypothesis, we could explore more precisely on the one hand the expansion of adjacent possible of neuroscience caused directly by AI, and on the other hand the expansion of the adjacent possible of other fields that have been influenced later by an adapted AI that was designed in neuroscience \citep{kauffman_investigations_2000,monechi_waves_2017,bianchini_artificial_2022}.

In addition to the limitations mentioned throughout this paper, especially for the building of our database described in Section~\ref{sec:data}, we should also consider diversity of types of publications that could differ from one discipline to another, and the manner they impact different disciplinary communities as well.
Indeed, we have shown that most of AI research published in peer-reviewed journals in neuroscience is impacting mainly neuroscience itself, but we could test whether this pattern subsists in other media for communicating research results, such as conference proceedings and preprints sharing platforms such as \textit{arXiv}, commonly used by mathematicians, physicists and computer scientists \citep{wainer_how_2013}.
Future research may clarify whether our results hold if we consider the impact of AI produced within neuroscience through these other publication outlets.

Finally, this paper intends to be a road map for further studies of the diffusion of AI in a broad range of disciplines or fields of research that are receptive to it, but probably with different patterns.
A comparative work would then be required.


\section{Declarations}

\begin{itemize}
    \item Ethics approval and consent to participate: Not applicable.
    \item Consent for publication: Not applicable.
    \item Availability of data, materials and codes: The datasets used for the this study are available at: \url{https://doi.org/10.5281/zenodo.10777508}.
    The codes for analyzing them are available at: \url{https://github.com/sysyMC/AI_in_Neuroscience_EpistemicIntegration_SocialSegregation}.
    \item Competing interests: The authors declare that they have no competing interests.
    \item Funding: S.F. is funded through a CNRS-MITI PhD grant within the project ``Epistemic Impact of Artificial Intelligence in Science'' (EpiAI). 
        This research has been partially supported by the ANR grant ScientIA (ANR-21-CE38-0020).
    \item Authors' contributions: S.F. collected and analyzed the data, discussed the results, wrote the manuscript.
        F.G. conceived the research, collected the data, discussed the results, wrote the manuscript.
        M.D. discussed the results, wrote the manuscript. 
        P.T. discussed the results, wrote the manuscript.
        All authors have read and approved the manuscript.
    \item Acknowledgements: We thank Abdelghani Maddi for very useful discussions.
\end{itemize}

\begin{appendices}

\section{Self-similarities of references and citations, and disciplinary concentration in the AI and non-AI corpora}
\label{app:citations_ranking_jaccard}

This section focuses on the cognitive development of the research field associated with the AI corpus ($\mathcal{P}\cap AI$) and the non-AI one ($\mathcal{P}\cap \overline{AI}$), based on the same similarity measure introduced in Section~\ref{sec:results_disc_jaccard}.
We compute for each corpus two such measures that could vary over time:
\begin{enumerate}
    \item the temporal self-similarity of one given ranking $r$, namely the similarity between this ranking at time $t$ and the the same at the previous time $t-1$, denoted as $J(r_{t-1}, r_t)$; we apply it for references' and citations' rankings in each corpus,
    \item the temporal disciplinary concentration of one corpus, namely the similarity between its references ranking $r_R$ and its citations one $r_I$ at a given time $t$, denoted as $J(r_R(t), r_I(t))$,
\end{enumerate}

\begin{figure}[htb!]
    \centering
    \includegraphics[width=0.9\textwidth]{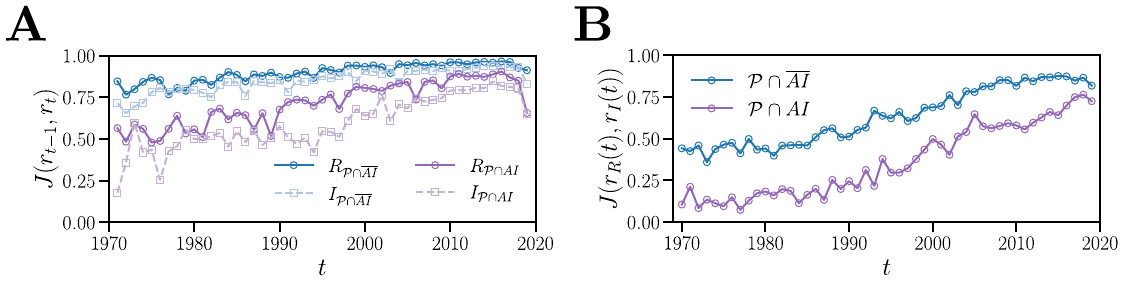}
    \caption{\textbf{A}: Temporal self-similarity of disciplinary rankings associated with either the references of one specific corpus ($R$) or the citations it has received ($I$), either for the AI-related corpus ($\mathcal{P}\cap AI$) or for the non-AI one ($\mathcal{P}\cap \overline{AI}$). 
    More precisely, one point at time $t$ is the similarity between the ranking at time $t$ and the former one at $t-1$. 
    \textbf{B}: Instantaneous similarity between the references used by one corpus and its produced citational impact.}
\label{fig:jaccard_app}
\end{figure}

The evolution of these two indices are represented in Figs.~\ref{fig:jaccard_app}A and \ref{fig:jaccard_app}B respectively.

According to Fig.~\ref{fig:jaccard_app}A, the references on which the AI corpus and the non-AI one draw upon respectively (solid lines) are both consolidating with time toward their highest respective values in 2017, but not at the same speed.
Indeed, the non-AI corpus lies on almost the same set of disciplines from year to year, while the AI-related one grow from a low similarity in 1970 (around 50\%) to a stable set with a high similarity (around 90\%) between 2015 and 2017.
The observed decrease after 2017 is mainly due to the lack of data grabbed by MAG.
In addition, while the self-similarity of citations in the corpus $\mathcal{P}\cap \overline{AI}$ (blue dashed line) follow the same trend as its corresponding self-similarity of references, the self-similarity of citations in the corpus $\mathcal{P}\cap \overline{AI}$ (purple dashed line) is rather chaotic and varying around 50\% between 1970 and the late 1990s, then it increases toward around 80\% until 2019.
This means that the short-term impacted fields in the two corpora are also consolidating toward rather the same ones in recent days.
More precisely, Fig.~\ref{fig:jaccard_app}B shows a growing concentration of references' and impact's rankings inside each corpus over the time period under study, ie. the impacted disciplines and those appearing in references become more and more similar, hence a research inspired by itself toward itself.
This shows also the progressive disciplinary homogenisation of the two corpora $\mathcal{P}\cap AI$ and $\mathcal{P}\cap \overline{AI}$ independently, and therefore a consolidation of the entire field of neuroscience.

\section{Disciplinary composition of the common interest area of the AI-related and non-AI corpora}
\label{app:disc_comp_cia}

\begin{figure}[htb!]
    \centering
    \includegraphics[width=0.8\textwidth]{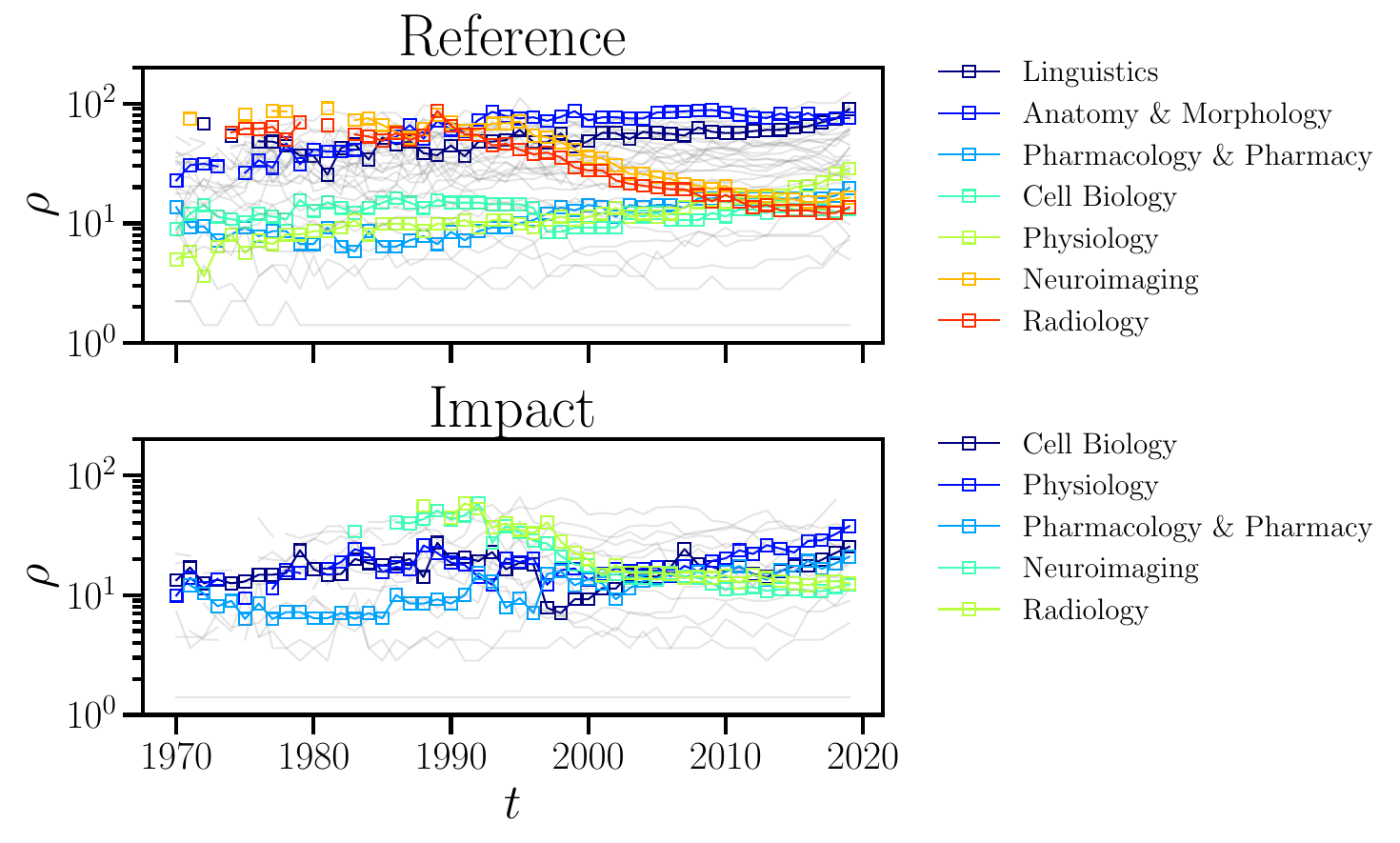}
    \caption{Time evolution of the distance $\rho$ of the disciplines included in the common interest area, as defined and shown in Fig.~\ref{fig:rankings-2D_plots}.
    Only the most significant curves are highlighted with colors.}
    \label{fig:rho_cia}
\end{figure}

This section focuses on the disciplines inside the common interest area of the AI and non-AI corpora defined in Section~\ref{sec:results_disc_carto}.
According to Fig.~\ref{fig:rankings-2D_plots} in the main text, each discipline $d$ appearing in the references and/or in the citations of these corpora is located in a 2D space by the coordinates ($r^d_{\mathcal{P}\cap AI},r^d_{\mathcal{P}\cap\overline{AI}}$) associated with its respective ranks in the AI and non-AI corpora.
From these coordinates we compute the distance of the disciplines from the origin of the map (point with ranks (0,0)), denoted as $\rho$ in the following.
Fig.~\ref{fig:rho_cia} shows the evolution of this distance for some disciplines, which are here exhibiting the most significant variations (increasing or decreasing) over the years since the 1970s.

This figure shows especially a spectacular rise since around 1995 of the references' rankings associated with \textit{Radiology} and \textit{Neuroimaging} in both AI and non-AI corpora, as well as a growing impact of neuroscience articles published in this period on these disciplines.
This trend thus testifies to the diffusion of these technologies in scientific and medical practices associated with neuroscience.

\section{Web of Science categories' abbreviations table}
\label{app:wos_abb_table}

\begin{tabular}{ll}
    \hline
    WOS Categories & Abbreviations \\
    \hline
    Acoustics & Acoustics \\
    Anatomy \& Morphology & A\&M \\
    Anesthesiology & Anesth \\
    Anthropology & Anthropo \\
    Automation \& Control Systems & A\&CS \\
    Biophysics & BioPhys \\
    Biochemical Research Method & BRM \\
    Biotechnology \& Applied Microbiology & Biotech \\
    Cardiac \& Cardiovascular System & Cardio \\
    Chemistry & Chem \\
    Computer Science & CS \\
    Critical Care Medicine & CCM \\
    Dentistry & Dentistry \\
    Developmental Biology & DB \\
    Ecology & Ecology \\
    Education & Educ \\
    Endocrinology \& Metabolism & E\&M \\
    Engineering & Engineering \\
    Entomology & Entomo \\
    Ergonomics & Ergo \\
    Gastroenterology \& Hepatology & G\&H \\
    Genetics \& Heredity & Genetics \\
    Geriatrics \& Gerontology & G\&G \\
    Hematology & Hemato \\
    Imaging Science \& Photographic Technology & IS\&PT \\
    Immunology & Immuno \\
    Instruments \& Instrumentation & Instrum \\
    Language \& Linguistics & L\&L \\
    Mathematical \& Computational Biology & M\&CB \\
    Mathematics & Maths \\
    Mechanics & Mech \\
    Medical Informatics & MI \\
    Medical Laboratory Technology & MLT \\
    Microbiology & Microbio \\
    Microscopy & Microscopy \\
    Neuroimaging & NI \\
    Nutrition \& Dietetics & N\&D \\
    Operations Research \& Management Science & OR\&MS \\
    Optics & Optics \\
    Otorhinolaryngology & Otorhino \\
    Pathology & Patho \\
    Pediatrics & Ped \\
    Peripheral Vascular Diseases & PVD \\
    Philosophy & Philo \\
    Physics & Phys \\
    Plant Sciences & Plant \\
    Public, Environmental \& Occupational Health & Public Health \\
    Radiology & Radio \\
    Rehabilitation & Rehab \\
    Social Sciences & SocSci \\
    Sport Sciences & Sport \\
    Statistics \& Probability & S\&P \\
    Substance Abuse & SA \\
    Telecommunications & Telecom \\
    Toxicology & Toxico \\
    Virology & Viro \\
    Zoology & Zoology \\
    \hline
\end{tabular}

\section{Temporal evolution of the inter-quartile collaboration network}
\label{app:collabNet_quartile_evol}

\begin{figure}[htb!]
    \centering
    \includegraphics[width=0.9\textwidth]{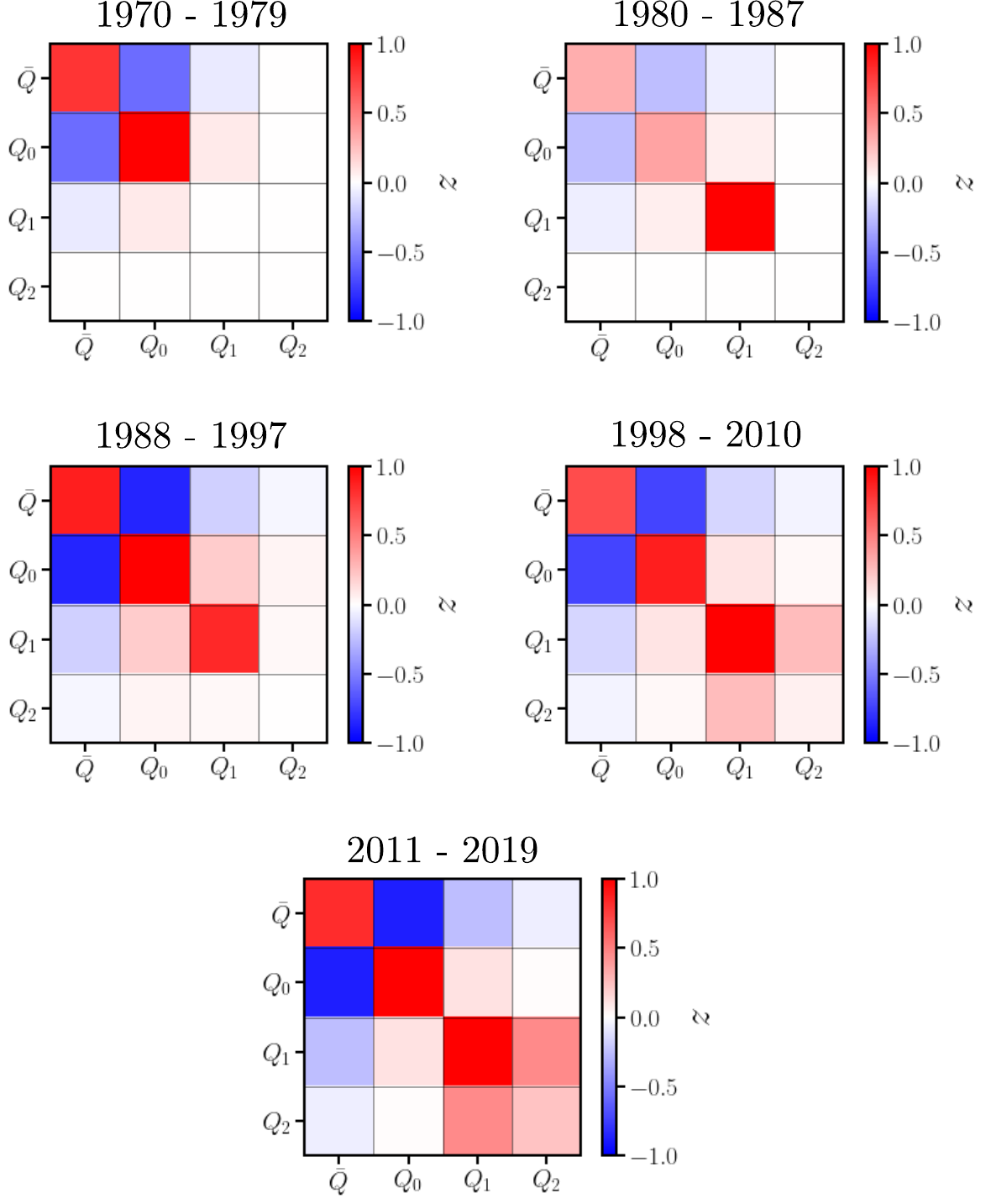}
    \caption{Evolution of the $z$-score matrix of the share of edges between $f_{AI}$ quartiles. 
    The first line, corresponding to the period 1970-1987, refers to the situation where the quartile $Q_2$ has not yet emerged in our dataset. 
    The separation of $\overline{Q}$ and the other quartiles $Q_i$'s occurs qualitatively in the period 1988-1997, when $Q_2$ was emerging and was beginning to publish in neuroscience. 
    The ending period spanning from 1998-2010, which concentrates 91\% of the 13,786,616 unweighted single edges in the temporal collaboration network TCN, is very similar to the configuration shown in the main text, especially in Fig.~\ref{fig:collabNet_quartile} which was computed on the basis of all the period 1970-2019.}
\label{fig:collabNet_quartile_evol}
\end{figure}

\end{appendices}

\bibliography{biblio}

\begin{thebibliography}{}
\renewcommand{\doi}[1]{\url{https://doi.org/#1}}
\bibcommenthead

\bibitem [\protect \citeauthoryear {%
Andler%
}{%
Andler%
}{%
{\protect \APACyear {1990}}%
}]{%
andler_connexionnisme_1990}
\APACinsertmetastar {%
andler_connexionnisme_1990}%
\begin{APACrefauthors}%
Andler, D.%
\end{APACrefauthors}%
\unskip\
\newblock
\APACrefYearMonthDay{1990}{}{}.
\newblock
{\BBOQ}\APACrefatitle {Connexionnisme et cognition: À la recherche des bonnes questions} {Connexionnisme et cognition: À la recherche des bonnes questions}.{\BBCQ}
\newblock
\APACjournalVolNumPages{Revue de Synth\`ese}{111}{1-2}{95--127,}
\newblock

\newblock

\PrintBackRefs{\CurrentBib}

\bibitem [\protect \citeauthoryear {%
Arencibia-Jorge%
, Vega-Almeida%
, Jiménez~Andrade%
\BCBL {}\ \BBA {} Carrillo-Calvet%
}{%
Arencibia-Jorge%
\ \protect \BOthers {.}}{%
{\protect \APACyear {2022}}%
}]{%
arencibia-jorge_evolution_2022}
\APACinsertmetastar {%
arencibia-jorge_evolution_2022}%
\begin{APACrefauthors}%
Arencibia-Jorge, R.%
, Vega-Almeida, R.L.%
, Jiménez~Andrade, J.L.%
\BCBL {} Carrillo-Calvet, H.%
\end{APACrefauthors}%
\unskip\
\newblock
\APACrefYearMonthDay{2022}{}{}.
\newblock
{\BBOQ}\APACrefatitle {Evolution and {Multidisciplinarity} of {Artificial} {Intelligence} {Research}} {Evolution and {Multidisciplinarity} of {Artificial} {Intelligence} {Research}}.{\BBCQ}
\newblock
\APACjournalVolNumPages{Scientometrics}{127}{2}{1--20,}
\newblock

\newblock

\PrintBackRefs{\CurrentBib}

\bibitem [\protect \citeauthoryear {%
Baruffaldi%
\ \protect \BOthers {.}}{%
Baruffaldi%
\ \protect \BOthers {.}}{%
{\protect \APACyear {2020}}%
}]{%
baruffaldi_identifying_2020}
\APACinsertmetastar {%
baruffaldi_identifying_2020}%
\begin{APACrefauthors}%
Baruffaldi, S.%
, van Beuzekom, B.%
, Dernis, H.%
, Harhoff, D.%
, Rao, N.%
, Rosenfeld, D.%
\BCBL {} Squicciarini, M.%
\end{APACrefauthors}%
\unskip\
\newblock
\APACrefYearMonthDay{2020}{}{}.
\newblock
\APACrefbtitle {Identifying and measuring developments in artificial intelligence: {Making} the impossible possible.} {Identifying and measuring developments in artificial intelligence: {Making} the impossible possible.}
\newblock
\APACrefnote{{OECD} {Science}, {Technology} and {Industry} {Working} {Papers}}
\PrintBackRefs{\CurrentBib}

\bibitem [\protect \citeauthoryear {%
Bianchini%
, Müller%
\BCBL {}\ \BBA {} Pelletier%
}{%
Bianchini%
\ \protect \BOthers {.}}{%
{\protect \APACyear {2022}}%
}]{%
bianchini_artificial_2022}
\APACinsertmetastar {%
bianchini_artificial_2022}%
\begin{APACrefauthors}%
Bianchini, S.%
, Müller, M.%
\BCBL {} Pelletier, P.%
\end{APACrefauthors}%
\unskip\
\newblock
\APACrefYearMonthDay{2022}{}{}.
\newblock
{\BBOQ}\APACrefatitle {Artificial intelligence in science: {An} emerging general method of invention} {Artificial intelligence in science: {An} emerging general method of invention}.{\BBCQ}
\newblock
\APACjournalVolNumPages{Research Policy}{51}{10}{,}
\newblock

\newblock

\PrintBackRefs{\CurrentBib}

\bibitem [\protect \citeauthoryear {%
Cardon%
, Cointet%
\BCBL {}\ \BBA {} Mazières%
}{%
Cardon%
\ \protect \BOthers {.}}{%
{\protect \APACyear {2018}}%
}]{%
cardon_neurons_2018}
\APACinsertmetastar {%
cardon_neurons_2018}%
\begin{APACrefauthors}%
Cardon, D.%
, Cointet, J\BHBI P.%
\BCBL {} Mazières, A.%
\end{APACrefauthors}%
\unskip\
\newblock
\APACrefYearMonthDay{2018}{}{}.
\newblock
{\BBOQ}\APACrefatitle {Neurons spike back. {The} invention of inductive machines and the artificial intelligence controversy} {Neurons spike back. {The} invention of inductive machines and the artificial intelligence controversy}{\BBCQ}\ (L.~Carey-Libbrecht, \BTRANS{}).
\newblock
\APACjournalVolNumPages{R\'eseaux}{211}{5}{173--220,}
\newblock

\newblock

\PrintBackRefs{\CurrentBib}

\bibitem [\protect \citeauthoryear {%
Cooper%
\ \BBA {} Shallice%
}{%
Cooper%
\ \BBA {} Shallice%
}{%
{\protect \APACyear {2010}}%
}]{%
cooper_cognitive_2010}
\APACinsertmetastar {%
cooper_cognitive_2010}%
\begin{APACrefauthors}%
Cooper, R.P.%
\BCBT {}\ \BBA {} Shallice, T.%
\end{APACrefauthors}%
\unskip\
\newblock
\APACrefYearMonthDay{2010}{}{}.
\newblock
{\BBOQ}\APACrefatitle {Cognitive {Neuroscience}: {The} {Troubled} {Marriage} of {Cognitive} {Science} and {Neuroscience}} {Cognitive {Neuroscience}: {The} {Troubled} {Marriage} of {Cognitive} {Science} and {Neuroscience}}.{\BBCQ}
\newblock
\APACjournalVolNumPages{Topics in Cognitive Science}{2}{3}{398--406,}
\newblock

\newblock

\PrintBackRefs{\CurrentBib}

\bibitem [\protect \citeauthoryear {%
Frank%
, Wang%
, Cebrian%
\BCBL {}\ \BBA {} Rahwan%
}{%
Frank%
\ \protect \BOthers {.}}{%
{\protect \APACyear {2019}}%
}]{%
frank_evolution_2019}
\APACinsertmetastar {%
frank_evolution_2019}%
\begin{APACrefauthors}%
Frank, M.R.%
, Wang, D.%
, Cebrian, M.%
\BCBL {} Rahwan, I.%
\end{APACrefauthors}%
\unskip\
\newblock
\APACrefYearMonthDay{2019}{}{}.
\newblock
{\BBOQ}\APACrefatitle {The evolution of citation graphs in artificial intelligence research} {The evolution of citation graphs in artificial intelligence research}.{\BBCQ}
\newblock
\APACjournalVolNumPages{Nature Machine Intelligence}{1}{2}{79--85,}
\newblock

\newblock

\PrintBackRefs{\CurrentBib}

\bibitem [\protect \citeauthoryear {%
Färber%
}{%
Färber%
}{%
{\protect \APACyear {2019}}%
}]{%
ghidini_microsoft_2019}
\APACinsertmetastar {%
ghidini_microsoft_2019}%
\begin{APACrefauthors}%
Färber, M.%
\end{APACrefauthors}%
\unskip\
\newblock
\APACrefYearMonthDay{2019}{}{}.
\newblock
{\BBOQ}\APACrefatitle {The {Microsoft} {Academic} {Knowledge} {Graph}: {A} {Linked} {Data} {Source} with 8 {Billion} {Triples} of {Scholarly} {Data}} {The {Microsoft} {Academic} {Knowledge} {Graph}: {A} {Linked} {Data} {Source} with 8 {Billion} {Triples} of {Scholarly} {Data}}.{\BBCQ}
\newblock
 C.~Ghidini\ \BOthers {.}\ (\BEDS), \APACrefbtitle {The {Semantic} {Web} – {ISWC} 2019} {The {Semantic} {Web} – {ISWC} 2019}\ (\BVOL\ 11779, \BPGS\ 113--129).
\newblock
\APACaddressPublisher{}{Springer International Publishing}.
\PrintBackRefs{\CurrentBib}

\bibitem [\protect \citeauthoryear {%
Färber%
}{%
Färber%
}{%
{\protect \APACyear {2020}}%
}]{%
mag_2020}
\APACinsertmetastar {%
mag_2020}%
\begin{APACrefauthors}%
Färber, M.%
\end{APACrefauthors}%
\unskip\
\newblock
\APACrefYearMonthDay{2020}{}{}.
\newblock
\APACrefbtitle {{Microsoft} {Academic} {Graph} records.} {{Microsoft} {Academic} {Graph} records.}
\newblock
\APACrefnote{Available at \url{https://zenodo.org/record/3936556}}
\PrintBackRefs{\CurrentBib}

\bibitem [\protect \citeauthoryear {%
Gao%
\ \BBA {} Wang%
}{%
Gao%
\ \BBA {} Wang%
}{%
{\protect \APACyear {2023}}%
}]{%
gao_quantifying_2023}
\APACinsertmetastar {%
gao_quantifying_2023}%
\begin{APACrefauthors}%
Gao, J.%
\BCBT {}\ \BBA {} Wang, D.%
\end{APACrefauthors}%
\unskip\
\newblock
\APACrefYearMonthDay{2023}{}{}.
\newblock
\APACrefbtitle {Quantifying the Benefit of Artificial Intelligence for Scientific Research.} {Quantifying the benefit of artificial intelligence for scientific research.}
\newblock
\APACrefnote{Preprint at \url{https://arxiv.org/abs/2304.10578}}
\PrintBackRefs{\CurrentBib}

\bibitem [\protect \citeauthoryear {%
Gargiulo%
, Caen%
, Lambiotte%
\BCBL {}\ \BBA {} Carletti%
}{%
Gargiulo%
\ \protect \BOthers {.}}{%
{\protect \APACyear {2016}}%
}]{%
gargiulo_classical_2016}
\APACinsertmetastar {%
gargiulo_classical_2016}%
\begin{APACrefauthors}%
Gargiulo, F.%
, Caen, A.%
, Lambiotte, R.%
\BCBL {} Carletti, T.%
\end{APACrefauthors}%
\unskip\
\newblock
\APACrefYearMonthDay{2016}{}{}.
\newblock
{\BBOQ}\APACrefatitle {The classical origin of modern mathematics} {The classical origin of modern mathematics}.{\BBCQ}
\newblock
\APACjournalVolNumPages{EPJ Data Science}{5}{1}{26,}
\newblock

\newblock

\PrintBackRefs{\CurrentBib}

\bibitem [\protect \citeauthoryear {%
Gargiulo%
, Fontaine%
, Dubois%
\BCBL {}\ \BBA {} Tubaro%
}{%
Gargiulo%
\ \protect \BOthers {.}}{%
{\protect \APACyear {2023}}%
}]{%
gargiulo_meso-scale_2023}
\APACinsertmetastar {%
gargiulo_meso-scale_2023}%
\begin{APACrefauthors}%
Gargiulo, F.%
, Fontaine, S.%
, Dubois, M.%
\BCBL {} Tubaro, P.%
\end{APACrefauthors}%
\unskip\
\newblock
\APACrefYearMonthDay{2023}{}{}.
\newblock
{\BBOQ}\APACrefatitle {A meso-scale cartography of the {AI} ecosystem} {A meso-scale cartography of the {AI} ecosystem}.{\BBCQ}
\newblock
\APACjournalVolNumPages{Quantitative Science Studies}{4}{3}{574--593,}
\newblock

\newblock

\PrintBackRefs{\CurrentBib}

\bibitem [\protect \citeauthoryear {%
Gopinath%
}{%
Gopinath%
}{%
{\protect \APACyear {2023}}%
}]{%
gopinath_artificial_2023}
\APACinsertmetastar {%
gopinath_artificial_2023}%
\begin{APACrefauthors}%
Gopinath, N.%
\end{APACrefauthors}%
\unskip\
\newblock
\APACrefYearMonthDay{2023}{}{}.
\newblock
{\BBOQ}\APACrefatitle {Artificial intelligence and neuroscience: {An} update on fascinating relationships} {Artificial intelligence and neuroscience: {An} update on fascinating relationships}.{\BBCQ}
\newblock
\APACjournalVolNumPages{Process Biochemistry}{125}{}{113--120,}
\newblock

\newblock

\PrintBackRefs{\CurrentBib}

\bibitem [\protect \citeauthoryear {%
Haas%
}{%
Haas%
}{%
{\protect \APACyear {1992}}%
}]{%
haas_introduction_1992}
\APACinsertmetastar {%
haas_introduction_1992}%
\begin{APACrefauthors}%
Haas, P.M.%
\end{APACrefauthors}%
\unskip\
\newblock
\APACrefYearMonthDay{1992}{}{}.
\newblock
{\BBOQ}\APACrefatitle {Introduction: {Epistemic} {Communities} and {International} {Policy} {Coordination}} {Introduction: {Epistemic} {Communities} and {International} {Policy} {Coordination}}.{\BBCQ}
\newblock
\APACjournalVolNumPages{International Organization}{46}{1}{1--35,}
\newblock

\newblock

\PrintBackRefs{\CurrentBib}

\bibitem [\protect \citeauthoryear {%
Hassabis%
, Kumaran%
, Summerfield%
\BCBL {}\ \BBA {} Botvinick%
}{%
Hassabis%
\ \protect \BOthers {.}}{%
{\protect \APACyear {2017}}%
}]{%
hassabis_neuroscience-inspired_2017}
\APACinsertmetastar {%
hassabis_neuroscience-inspired_2017}%
\begin{APACrefauthors}%
Hassabis, D.%
, Kumaran, D.%
, Summerfield, C.%
\BCBL {} Botvinick, M.%
\end{APACrefauthors}%
\unskip\
\newblock
\APACrefYearMonthDay{2017}{}{}.
\newblock
{\BBOQ}\APACrefatitle {Neuroscience-{Inspired} {Artificial} {Intelligence}} {Neuroscience-{Inspired} {Artificial} {Intelligence}}.{\BBCQ}
\newblock
\APACjournalVolNumPages{Neuron}{95}{2}{245--258,}
\newblock

\newblock

\PrintBackRefs{\CurrentBib}

\bibitem [\protect \citeauthoryear {%
Hentschel%
}{%
Hentschel%
}{%
{\protect \APACyear {2015}}%
}]{%
hentschel_periodization_2015}
\APACinsertmetastar {%
hentschel_periodization_2015}%
\begin{APACrefauthors}%
Hentschel, K.%
\end{APACrefauthors}%
\unskip\
\newblock
\APACrefYearMonthDay{2015}{}{}.
\newblock
{\BBOQ}\APACrefatitle {A periodization of research technologies and of the emergency of genericity} {A periodization of research technologies and of the emergency of genericity}.{\BBCQ}
\newblock
\APACjournalVolNumPages{Studies in History and Philosophy of Modern Physics}{52}{}{223--233,}
\newblock

\newblock

\PrintBackRefs{\CurrentBib}

\bibitem [\protect \citeauthoryear {%
Kauffman%
}{%
Kauffman%
}{%
{\protect \APACyear {2000}}%
}]{%
kauffman_investigations_2000}
\APACinsertmetastar {%
kauffman_investigations_2000}%
\begin{APACrefauthors}%
Kauffman, S.A.%
\end{APACrefauthors}%
\unskip\
\newblock
\APACrefYear{2000}.
\newblock
\APACrefbtitle {Investigations} {Investigations}.
\newblock
\APACaddressPublisher{}{Oxford University Press}.
\PrintBackRefs{\CurrentBib}

\bibitem [\protect \citeauthoryear {%
Lake%
, Ullman%
, Tenenbaum%
\BCBL {}\ \BBA {} Gershman%
}{%
Lake%
\ \protect \BOthers {.}}{%
{\protect \APACyear {2017}}%
}]{%
lake_building_2017}
\APACinsertmetastar {%
lake_building_2017}%
\begin{APACrefauthors}%
Lake, B.M.%
, Ullman, T.D.%
, Tenenbaum, J.B.%
\BCBL {} Gershman, S.J.%
\end{APACrefauthors}%
\unskip\
\newblock
\APACrefYearMonthDay{2017}{}{}.
\newblock
{\BBOQ}\APACrefatitle {Building {Machines} {That} {Learn} and {Think} {Like} {People}} {Building {Machines} {That} {Learn} and {Think} {Like} {People}}.{\BBCQ}
\newblock
\APACjournalVolNumPages{Behavioral and Brain Sciences}{40}{e253}{1--72,}
\newblock

\newblock

\PrintBackRefs{\CurrentBib}

\bibitem [\protect \citeauthoryear {%
Liu%
, Shapira%
\BCBL {}\ \BBA {} Yue%
}{%
Liu%
\ \protect \BOthers {.}}{%
{\protect \APACyear {2021}}%
}]{%
liu_tracking_2021}
\APACinsertmetastar {%
liu_tracking_2021}%
\begin{APACrefauthors}%
Liu, N.%
, Shapira, P.%
\BCBL {} Yue, X.%
\end{APACrefauthors}%
\unskip\
\newblock
\APACrefYearMonthDay{2021}{}{}.
\newblock
{\BBOQ}\APACrefatitle {Tracking developments in artificial intelligence research: constructing and applying a new search strategy} {Tracking developments in artificial intelligence research: constructing and applying a new search strategy}.{\BBCQ}
\newblock
\APACjournalVolNumPages{Scientometrics}{126}{4}{3153--3192,}
\newblock

\newblock

\PrintBackRefs{\CurrentBib}

\bibitem [\protect \citeauthoryear {%
Marcovich%
\ \BBA {} Shinn%
}{%
Marcovich%
\ \BBA {} Shinn%
}{%
{\protect \APACyear {2020}}%
}]{%
marcovich_science_2020}
\APACinsertmetastar {%
marcovich_science_2020}%
\begin{APACrefauthors}%
Marcovich, A.%
\BCBT {}\ \BBA {} Shinn, T.%
\end{APACrefauthors}%
\unskip\
\newblock
\APACrefYearMonthDay{2020}{}{}.
\newblock
{\BBOQ}\APACrefatitle {Science research regimes as architectures of knowledge in context: {A} ‘longue durée’ comparative historical sociology of structures and dynamics in science} {Science research regimes as architectures of knowledge in context: {A} ‘longue durée’ comparative historical sociology of structures and dynamics in science}.{\BBCQ}
\newblock
\APACjournalVolNumPages{Social Science Information}{59}{2}{310--328,}
\newblock

\newblock

\PrintBackRefs{\CurrentBib}

\bibitem [\protect \citeauthoryear {%
McCarthy%
}{%
McCarthy%
}{%
{\protect \APACyear {1981}}%
}]{%
mccarthy_epistemological_1981}
\APACinsertmetastar {%
mccarthy_epistemological_1981}%
\begin{APACrefauthors}%
McCarthy, J.%
\end{APACrefauthors}%
\unskip\
\newblock
\APACrefYearMonthDay{1981}{}{}.
\newblock
{\BBOQ}\APACrefatitle {Epistemological {Problems} of {Artificial} {Intelligence}} {Epistemological {Problems} of {Artificial} {Intelligence}}.{\BBCQ}
\newblock
 \APACrefbtitle {Readings in {Artificial} {Intelligence}} {Readings in {Artificial} {Intelligence}}\ (\BPGS\ 459--465).
\newblock
\APACaddressPublisher{}{Elsevier}.
\PrintBackRefs{\CurrentBib}

\bibitem [\protect \citeauthoryear {%
Monechi%
, Ruiz-Serrano%
, Tria%
\BCBL {}\ \BBA {} Loreto%
}{%
Monechi%
\ \protect \BOthers {.}}{%
{\protect \APACyear {2017}}%
}]{%
monechi_waves_2017}
\APACinsertmetastar {%
monechi_waves_2017}%
\begin{APACrefauthors}%
Monechi, B.%
, Ruiz-Serrano, {\~A}.%
, Tria, F.%
\BCBL {} Loreto, V.%
\end{APACrefauthors}%
\unskip\
\newblock
\APACrefYearMonthDay{2017}{}{}.
\newblock
{\BBOQ}\APACrefatitle {Waves of novelties in the expansion into the adjacent possible} {Waves of novelties in the expansion into the adjacent possible}.{\BBCQ}
\newblock
\APACjournalVolNumPages{Plos One}{12}{6}{,}
\newblock

\newblock

\PrintBackRefs{\CurrentBib}

\bibitem [\protect \citeauthoryear {%
Perconti%
\ \BBA {} Plebe%
}{%
Perconti%
\ \BBA {} Plebe%
}{%
{\protect \APACyear {2020}}%
}]{%
perconti_deep_2020}
\APACinsertmetastar {%
perconti_deep_2020}%
\begin{APACrefauthors}%
Perconti, P.%
\BCBT {}\ \BBA {} Plebe, A.%
\end{APACrefauthors}%
\unskip\
\newblock
\APACrefYearMonthDay{2020}{}{}.
\newblock
{\BBOQ}\APACrefatitle {Deep learning and cognitive science} {Deep learning and cognitive science}.{\BBCQ}
\newblock
\APACjournalVolNumPages{Cognition}{203}{}{12,}
\newblock

\newblock

\PrintBackRefs{\CurrentBib}

\bibitem [\protect \citeauthoryear {%
Roth%
}{%
Roth%
}{%
{\protect \APACyear {2008}}%
}]{%
roth_reseaux_2008}
\APACinsertmetastar {%
roth_reseaux_2008}%
\begin{APACrefauthors}%
Roth, C.%
\end{APACrefauthors}%
\unskip\
\newblock
\APACrefYearMonthDay{2008}{}{}.
\newblock
{\BBOQ}\APACrefatitle {Réseaux épistémiques : formaliser la cognition distribuée} {Réseaux épistémiques : formaliser la cognition distribuée}.{\BBCQ}
\newblock
\APACjournalVolNumPages{Sociologie du Travail}{50}{3}{353--371,}
\newblock

\newblock

\PrintBackRefs{\CurrentBib}

\bibitem [\protect \citeauthoryear {%
Savage%
}{%
Savage%
}{%
{\protect \APACyear {2019}}%
}]{%
savage_how_2019}
\APACinsertmetastar {%
savage_how_2019}%
\begin{APACrefauthors}%
Savage, N.%
\end{APACrefauthors}%
\unskip\
\newblock
\APACrefYearMonthDay{2019}{}{}.
\newblock
{\BBOQ}\APACrefatitle {How {AI} and neuroscience drive each other forwards} {How {AI} and neuroscience drive each other forwards}.{\BBCQ}
\newblock
\APACjournalVolNumPages{Nature}{571}{7766}{15--17,}
\newblock

\newblock

\PrintBackRefs{\CurrentBib}

\bibitem [\protect \citeauthoryear {%
Schuchmann%
}{%
Schuchmann%
}{%
{\protect \APACyear {2019}}%
}]{%
schuchmann_history_2019}
\APACinsertmetastar {%
schuchmann_history_2019}%
\begin{APACrefauthors}%
Schuchmann, S.%
\end{APACrefauthors}%
\unskip\
\newblock
\APACrefYearMonthDay{2019}{}{}.
\newblock
\APACrefbtitle {History of the {Second} {AI} {Winter}.} {History of the {Second} {AI} {Winter}.}
\newblock
\APACrefnote{Medium. \url{https://towardsdatascience.com/history-of-the-second-ai-winter-406f18789d45} (visited on 2023-04-13)}
\PrintBackRefs{\CurrentBib}

\bibitem [\protect \citeauthoryear {%
Sedooka%
, Steffen%
, Paulsen%
\BCBL {}\ \BBA {} Darbellay%
}{%
Sedooka%
\ \protect \BOthers {.}}{%
{\protect \APACyear {2015}}%
}]{%
sedooka_paradoxe_2015}
\APACinsertmetastar {%
sedooka_paradoxe_2015}%
\begin{APACrefauthors}%
Sedooka, A.%
, Steffen, G.%
, Paulsen, T.%
\BCBL {} Darbellay, F.%
\end{APACrefauthors}%
\unskip\
\newblock
\APACrefYearMonthDay{2015}{}{}.
\newblock
{\BBOQ}\APACrefatitle {Paradoxe identitaire et interdisciplinarité : un regard sur les identités disciplinaires des chercheurs} {Paradoxe identitaire et interdisciplinarité : un regard sur les identités disciplinaires des chercheurs}.{\BBCQ}
\newblock
\APACjournalVolNumPages{Natures Sciences Sociétés}{23}{4}{367--377,}
\newblock

\newblock

\PrintBackRefs{\CurrentBib}

\bibitem [\protect \citeauthoryear {%
Shinn%
\ \BBA {} Joerges%
}{%
Shinn%
\ \BBA {} Joerges%
}{%
{\protect \APACyear {2002}}%
}]{%
shinn_transverse_2002}
\APACinsertmetastar {%
shinn_transverse_2002}%
\begin{APACrefauthors}%
Shinn, T.%
\BCBT {}\ \BBA {} Joerges, B.%
\end{APACrefauthors}%
\unskip\
\newblock
\APACrefYearMonthDay{2002}{}{}.
\newblock
{\BBOQ}\APACrefatitle {The {Transverse} {Science} and {Technology} {Culture}: {Dynamics} and {Roles} of {Research}-{Technology}} {The {Transverse} {Science} and {Technology} {Culture}: {Dynamics} and {Roles} of {Research}-{Technology}}.{\BBCQ}
\newblock
\APACjournalVolNumPages{Social Science Information}{41}{2}{207--251,}
\newblock

\newblock

\PrintBackRefs{\CurrentBib}

\bibitem [\protect \citeauthoryear {%
Tang%
, Li%
\BCBL {}\ \BBA {} Ma%
}{%
Tang%
\ \protect \BOthers {.}}{%
{\protect \APACyear {2022}}%
}]{%
tang_internationalizing_2022}
\APACinsertmetastar {%
tang_internationalizing_2022}%
\begin{APACrefauthors}%
Tang, X.%
, Li, X.%
\BCBL {} Ma, F.%
\end{APACrefauthors}%
\unskip\
\newblock
\APACrefYearMonthDay{2022}{}{}.
\newblock
{\BBOQ}\APACrefatitle {Internationalizing {AI}: evolution and impact of distance factors} {Internationalizing {AI}: evolution and impact of distance factors}.{\BBCQ}
\newblock
\APACjournalVolNumPages{Scientometrics}{127}{1}{181--205,}
\newblock

\newblock

\PrintBackRefs{\CurrentBib}

\bibitem [\protect \citeauthoryear {%
Visser%
, Van~Eck%
\BCBL {}\ \BBA {} Waltman%
}{%
Visser%
\ \protect \BOthers {.}}{%
{\protect \APACyear {2021}}%
}]{%
visser_comparison_2021}
\APACinsertmetastar {%
visser_comparison_2021}%
\begin{APACrefauthors}%
Visser, M.%
, Van~Eck, N.J.%
\BCBL {} Waltman, L.%
\end{APACrefauthors}%
\unskip\
\newblock
\APACrefYearMonthDay{2021}{}{}.
\newblock
{\BBOQ}\APACrefatitle {Large-scale comparison of bibliographic data sources: {Scopus}, {Web} of {Science}, {Dimensions}, {Crossref}, and {Microsoft} {Academic}} {Large-scale comparison of bibliographic data sources: {Scopus}, {Web} of {Science}, {Dimensions}, {Crossref}, and {Microsoft} {Academic}}.{\BBCQ}
\newblock
\APACjournalVolNumPages{Quantitative Science Studies}{2}{1}{20--41,}
\newblock

\newblock

\PrintBackRefs{\CurrentBib}

\bibitem [\protect \citeauthoryear {%
Wainer%
, Eckmann%
, Goldenstein%
\BCBL {}\ \BBA {} Rocha%
}{%
Wainer%
\ \protect \BOthers {.}}{%
{\protect \APACyear {2013}}%
}]{%
wainer_how_2013}
\APACinsertmetastar {%
wainer_how_2013}%
\begin{APACrefauthors}%
Wainer, J.%
, Eckmann, M.%
, Goldenstein, S.%
\BCBL {} Rocha, A.%
\end{APACrefauthors}%
\unskip\
\newblock
\APACrefYearMonthDay{2013}{}{}.
\newblock
{\BBOQ}\APACrefatitle {How productivity and impact differ across computer science subareas} {How productivity and impact differ across computer science subareas}.{\BBCQ}
\newblock
\APACjournalVolNumPages{Communications of the ACM}{56}{8}{67--73,}
\newblock

\newblock

\PrintBackRefs{\CurrentBib}

\end{thebibliography}

\end{document}